\documentclass[journal]{IEEEtran}
\ifCLASSINFOpdf

\else

\fi

\usepackage{microtype}
\usepackage{graphicx}
\usepackage{booktabs} 
\usepackage{amsmath,amssymb}
\usepackage{enumerate}
\usepackage{subcaption}
\usepackage[linesnumbered,ruled,vlined]{algorithm2e}
\usepackage{algorithmic}
\usepackage[hidelinks]{hyperref}
\usepackage{mathrsfs}
\usepackage{xcolor}
\usepackage{booktabs}
 \usepackage{todonotes}
\usepackage{cite}
\usepackage{dsfont}

\usepackage{float}

\usepackage[normalem]{ulem}

\usepackage{amsthm} 

\newcommand{\bd}{\mathbf}

\newcommand{\solar}{\mathdutchcal{s}}

\definecolor{dartmouthgreen}{rgb}{0.05, 0.5, 0.06}
\definecolor{bittersweet}{rgb}{1.0, 0.44, 0.37}
\definecolor{brightmaroon}{rgb}{0.76, 0.13, 0.28}
\definecolor{bluegray}{rgb}{0.4, 0.6, 0.8}
\definecolor{brandeisblue}{rgb}{0.0, 0.44, 1.0}

\definecolor{amethyst}{rgb}{0.6, 0.4, 0.8}
\newcommand\amberout{\bgroup\markoverwith{\textcolor{amethyst}{\rule[0.5ex]{2pt}{1pt}}}\ULon}

\definecolor{blue(ryb)}{rgb}{0.01, 0.28, 1.0}
\definecolor{brickred}{rgb}{0.8, 0.25, 0.33}
\newcommand\aidanout{\bgroup\markoverwith{\textcolor{brickred}{\rule[0.5ex]{2pt}{1pt}}}\ULon}

\DeclareMathAlphabet{\mathdutchcal}{U}{dutchcal}{m}{n}
\SetMathAlphabet{\mathdutchcal}{bold}{U}{dutchcal}{b}{n}
\DeclareMathAlphabet{\mathdutchbcal}{U}{dutchcal}{b}{n}

\begin{document}

\title{Operating Dynamic Reserve Dimensioning Using Probabilistic Forecasts}

%
%
%

\author{Napoleon~Costilla-Enriquez\!\textsuperscript{*}\thanks{\textsuperscript{*}Authors contributed equally.},
        ~Miguel~Ortega-Vazquez\!\textsuperscript{*},
        ~Aidan~Tuohy,
        ~Amber~Motley, and
        ~Rebecca~Webb.
}

\maketitle

\begin{abstract}
The rapid integration of variable energy sources (VRES) into power grids increases variability and uncertainty of the net demand, making the power system operation challenging. Operating reserve is used by system operators to manage and hedge against such variability and uncertainty.  Traditionally, reserve requirements are determined by rules-of-thumb (static reserve requirements, e.g., NERC Reliability Standards), and more recently, dynamic reserve requirements from tools and methods which are in the adoption process (e.g., DynADOR, DRD, and RESERVE, among others). While these methods/tools significantly improve the static rule-of-thumb approaches, they rely exclusively on deterministic data (i.e., best guess only). Consequently, these methods disregard the probabilistic uncertainty thresholds associated with specific days and their weather conditions (i.e., best guess plus probabilistic uncertainty).

This work presents practical approaches to determine the operating reserve requirements leveraging the wealth information from probabilistic forecasts. 
Proposed approaches are validated and tested using actual data from the CAISO system.  
Results show the benefits in terms of risk reduction of considering the probabilistic forecast information into the dimensioning process of operating reserve requirements.

\end{abstract}

\begin{IEEEkeywords}
Operating reserve, probabilistic forecasts, uncertainty, variable renewable energy sources, generation scheduling.
\end{IEEEkeywords}

\IEEEpeerreviewmaketitle

\section{Introduction}
The need to curb emissions from the electric sector has driven the integration of variable renewable energy sources (VRES) in power systems.  The penetration of VRES is expected to increase as emission targets are revisited and enforced in various countries, e.g.,~\cite{briefingroom2021}.  Being weather-driven, the production from VRES is variable and uncertain.  In order to accommodate these characteristics, system operators need to allocate operating reserves\footnote{In this work, operating reserve is defined as the headroom and footroom held a scheduling process, e.g., day-ahead; and released in a subsequent process, e.g., real-time, to accommodate deviations from forecasts.} to ensure that power balance is maintained regardless of the materialized VRES production. 
Currently, power system operating plans are determined using deterministic forecasts.  That is, each of the variables that are subject to uncertainty (e.g., load, wind, and solar) is represented by a single time series over the decision horizon. 
The expected uncertainty is then estimated via regression analyses of equivalent operating conditions, e.g., ~\cite{bouffard2011value,dvorkin2014assessing,ortega2008estimating,ela2011operating,holttinen2012methodologies}.

While practical, such approaches rely only on the most likely trajectory of each stochastic variable with an added headroom and footroom from the regression analysis but disregard the likelihood and extent of the different uncertainty thresholds.  Similarly, these approaches fail to capture specific or unique characteristics of the day being planned for (e.g., expected clouds at a particular day hour not captured in the regression analysis).  
In contrast, probabilistic forecasts take the form of a predictive probability distribution over the decision horizon of the quantities of interest (e.g., solar generation),~\cite{hong2020energy}.  Such probability distributions furnish the user with a wealth of additional information on the upper and lower bounds of different probabilistic thresholds, Fig.~\ref{fig:probabilistic_forecast}.  Having access to such information allows system operators to pre-position the scheduling and dispatch to respond to uncertainty taking into account all its attributes (e.g., mean, variance, skewness, and kurtosis).  For instance, if the uncertain envelope for solar production at a particular time period is narrow and symmetric, the upward and downward reserve requirements would be modest and of equivalent magnitude.  If, on the other hand, the solar probabilistic forecast is wide and skewed toward low realizations, then the operator should schedule larger amounts of upward reserves and not so much downward reserves.

Probabilistic forecasts are starting to be used by utilities and system operators to hedge their operating plans against uncertainty better.  There are a number of initiatives targeting various objectives, including scheduling process, operational planning, enhanced forecasting approaches, and energy trading~\cite{li2020review}, among others.
In~\cite{krishnan2019ramping}, the authors enhance different methods to forecast better ramping products and regulation reserves using probabilistic solar power forecasts.  In~\cite{tuohy2019operational} the authors propose different methods to determine reserves to respond to uncertainty and assess the operating reserve requirements using production cost model simulations.  This manuscript is a result of such work. 
The authors in~\cite{etingov2018balancing} predict requirements for regulation and load following requirements harnessing probabilistic forecast information. 
These different efforts have been proposed for renewable energy, starting with wind power forecasts and more recently solar power~\cite{pinson2012very,trombe2012general,anastasiades2013quantile}.  Much of the work has focused on methods to develop better probabilistic forecasts, and only recently has it moved to applications.

The aim of these and other ongoing works is to produce systematic algorithms to further integrate probabilistic forecasts into utilities and ISOs workflows~\cite{haupt2019use}. 
To date, utilities and ISOs do not explicitly utilize probabilistic forecasts in the scheduling and dispatch stages. 
To bridge this gap, this work proposes different recursive and anticipative models to account for probabilistic uncertainty explicitly in the dynamic operating reserve determination.  The resulting probabilistic dynamic reserve requirements reduce the risk of undergoing recorded deviations or even hedge against anticipated deviations captured in the probabilistic forecast (e.g., particular weather-driven events for the period under study).  The probabilistic reserve requirements are tested using data from CAISO (California Independent System Operator) and compared against a baseline approach (reserve determination with deterministic forecasts).   

The paper is organized as follows. Section~\ref{sec:PF_scenarios} presents a method to convey probabilistic information via probabilistic scenarios, Section~\ref{sec:dynamic_reserve} introduces a method to determine dynamic reserve requirements,  Section~\ref{sec:probabilistic_reserve} proposes different approaches to determine reserve requirements using probabilistic forecasts.  Section~\ref{sec:results} presents numerical experiments and results, and Section~\ref{sec:conclusion} concludes the paper.

\section{Transformation of Probabilistic Information}\label{sec:PF_scenarios}

Probabilistic forecasts are more informative than deterministic forecasts, but they cannot be used directly in existing scheduling or actual operation models (e.g., PLEXOS~\cite{plexos}, PSO~\cite{PSO}, PROMOD~\cite{PROMOD}, MAPS~\cite{MAPS}).
As a consequence, the probabilistic information needs to be transformed into a dataset that contains the same information but in a usable format.  In this case, the probabilistic information is transformed into \emph{probabilistic scenarios}, to be used in tools to determine operating reserves (e.g., EPRI's DynADOR\footnote{To the best of authors' knowledge, DynADOR is the only reserve determination tool used by utilities and ISOs that can incorporate probabilistic forecasts directly.}~\cite{dynador}, N-SIDE's DRD~\cite{n_side}, E3's RESERVE~\cite{E3}), or even advanced scheduling platforms (e.g., PLEXOS and PSO). 
%
\begin{figure}[htb]
 \centering
 \begin{subfigure}[t]{1.72in}
     \centering
     \includegraphics[width=\textwidth]{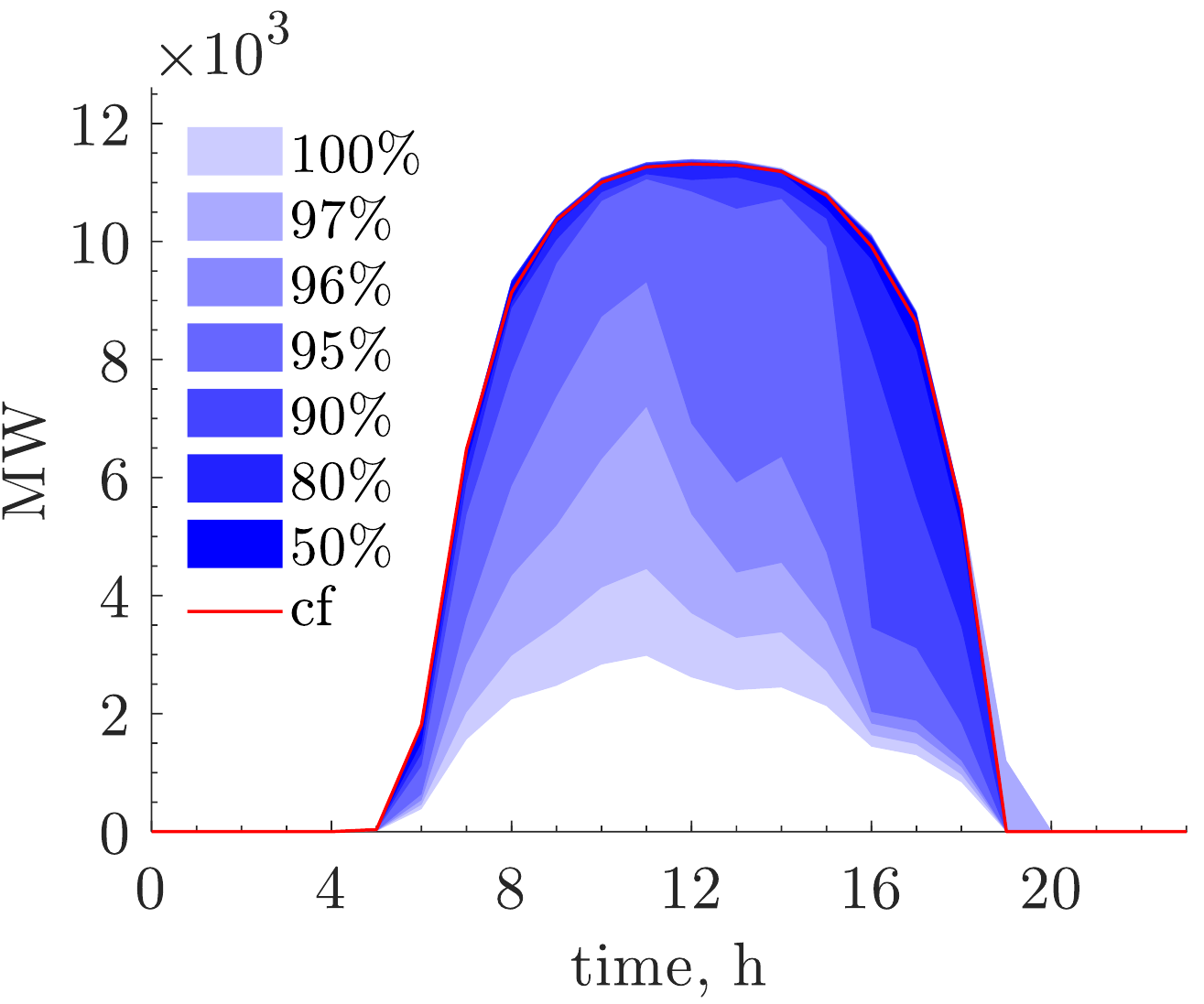}
     \caption{Probabilistic forecast}
     \label{fig:probabilistic_forecast}
 \end{subfigure}
 \hfill
 \begin{subfigure}[t]{1.72in}
     \centering
     \includegraphics[width=\textwidth]{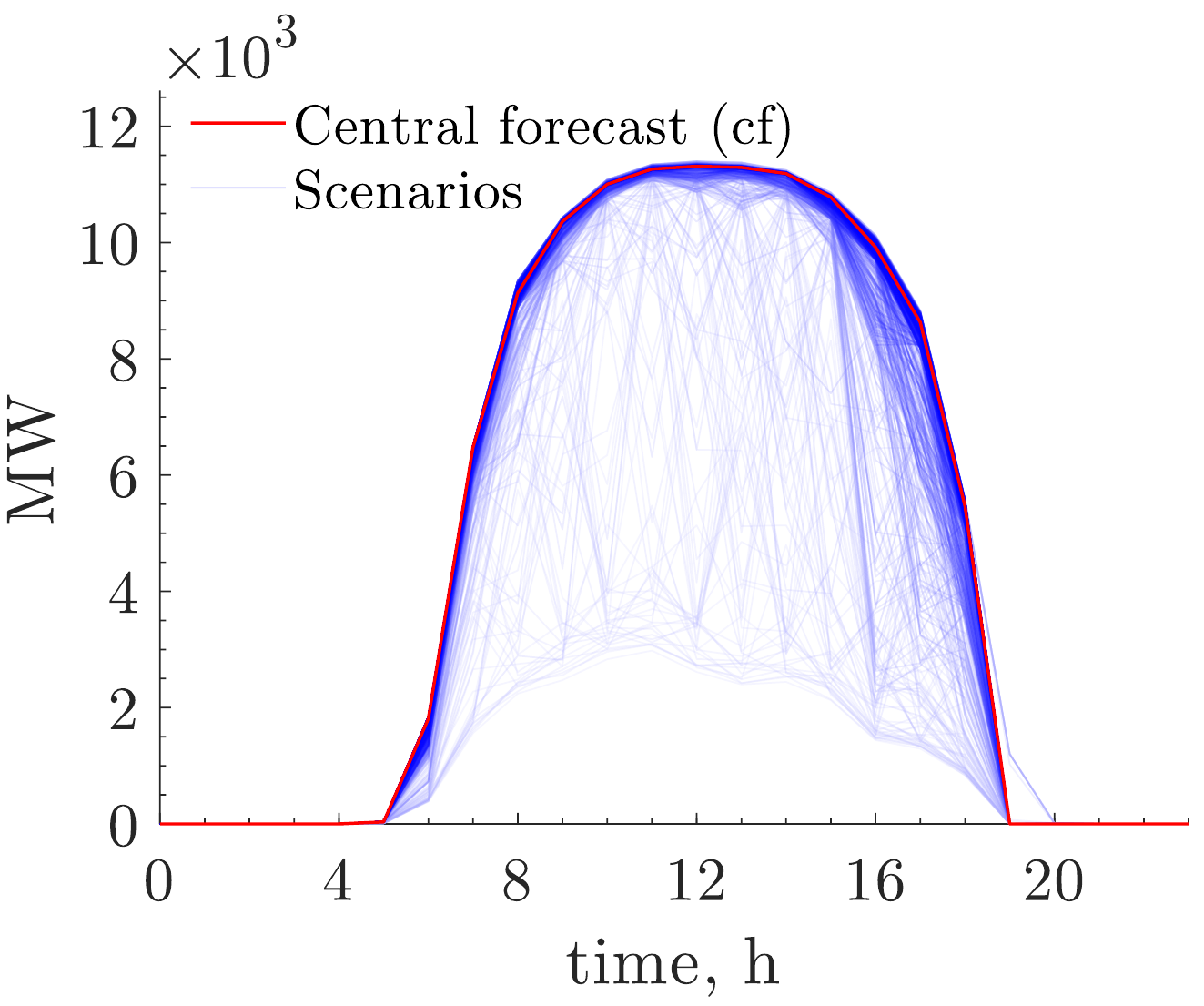}
     \caption{Probabilistic scenarios}
     \label{fig:probabilistic_scenarios}
 \end{subfigure}
    \caption{Probabilistic forecast and probabilistic scenarios for aggregated solar generation in CAISO on July $1^{\rm st}$, 2019.}
    \label{fig:forecast_scenario}
\end{figure}

To be able to use the probabilistic information, Fig.~\ref{fig:probabilistic_forecast}, in reserve determination tools or even production cost models, they first have to be transformed from probabilistic thresholds to chronologically correlated probabilistic scenarios, Fig.~\ref{fig:probabilistic_scenarios}.  This is done by following the steps outlined below.
\begin{enumerate}[(i)]
\item{Generation of random variates:} In this step $S$ vectors of length $T$ of standard normally distributed samples are produced, each of them being of the form: $\mathbf{x}_s = \left(x_1, x_2, \dots, x_T\right)$.  Following a standard normal distribution implies that each sample in the vector has zero mean (i.e., no bias) and a standard deviation equal to 1 (i.e., $x_t \sim \mathcal{N}(0,1)$).

\item {Induce inter-temporal correlation:} To ensure coherence among periods over the horizon $T$, it is essential to model the correlation that exists among samples in $\mathbf{x}$.  To this end, the covariance matrix $\mathbf{\Sigma}$ is constructed and it is expressed as a function of the temporal correlation coefficients ${\theta}$ and correlation decay $\mathbf{\omega}$, as $\mathbf{\Sigma} = \boldsymbol{\sigma} \boldsymbol{\theta} \boldsymbol{\sigma}$, where $\boldsymbol{\sigma}$ is equal to the identity matrix  $\bd{I}$ and therefore: 
\begin{equation}\label{eq:correlation}
\mathbf{\Sigma}\!=\!\!\! 
\begin{bmatrix}
1                & \theta  & \theta\!-\!\omega & \ldots  & \theta\!-\!(T\!-\!2)\omega \\
\theta             & 1     & \theta        &         &                 \\
\vdots           &      &              & \ddots  &                 \\
\theta\!-\!(T\!-\!2)\omega &      &              &        &       1
\end{bmatrix}\!\!.
\end{equation}

Note that in equation~\eqref{eq:correlation}, the correlation decay is modeled as a linear function of time.  However, the decay could drop at a different rate.  This information is obtained recursively from historical data~\cite{pinson2009probabilistic}.
Being symmetric, real-valued, and positive definite, the covariance matrix can be decomposed using its Cholesky factor:
\begin{equation}
\mathbf{\Sigma} = \mathbf{C \, C^{\rm T}},
\end{equation}
where superscript `${\rm T}$' denotes the transpose of the matrix. Pre-multiplying $\mathbf{x}$ by $\mathbf{C}$  gives the desired vector of correlated variates $\mathbf{y}$,~\cite{ortega2010assessing}:
\begin{equation}
\mathbf{y}_s = \mathbf{C} \, \mathbf{x}_s .
\end{equation}

\item {Induce probabilistic forecast distributions:} The vectors generated in step (ii) follow normal distributions,~\cite{kay2006intuitive}, which is not the case in probabilistic forecasts.  To transform the chronologically-correlated standard normally distributed variates to follow the probabilistic forecast CDFs, it is first required to pass each sample of $\mathbf{y}_s$ through the inverse probit function ($\Phi(\cdot)$),~\cite{devore1987probability}:
\begin{equation}
\mathbf{z}_{s} = \Phi(\mathbf{y}_{s}).
\end{equation}
The vector $\mathbf{z}_s$ is then a set of temporally correlated uniform variates in the range $[0,1]$.  Each sample in $\mathbf{z}_s$ is then passed through the chronologically corresponding non-parametric CDF ($F_t(\cdot)$) of the probabilistic forecast, to then obtain the desired scenario ``$\mathbf{s}$'':
\begin{equation}
s_t = F_t(z_{t}), \quad t=1,\dots,T.
\end{equation}

Steps (i)-(iii) are repeated as many times as scenarios `$S$' are required to produce a statistically sufficient population that matches the mathematical moments of the probabilistic forecast.  Details of this process, as well as reduction techniques to maintain statistical properties are available in~\cite{bhavsar2021machine}.  

\item {Scenarios' probabilities:}  Given the location of each time sample of the scenario in the uncertainty envelope, they would have different overall probabilities of materializing.  The probability of each scenario is then given as:   
%
\begin{equation}
\pi_s = \prod_{t=1}^T P_{t}^{(b)}, \quad  s=1,\dots, S,
\end{equation}
where `$b$' indicates the threshold of the probabilistic forecast where the sample materialized, and $P_{t}^{(b)}$ its probability (e.g., see ~\cite{ortega2008estimating}).
Note that only a finite number of scenarios $S$ is generated to form a set of scenarios $\bd{S} = \left\{\bd s_1, \bd s_2, \dots, \bd s_{S}\right\}$ associated with their respective probabilities $\boldsymbol{\pi} = \left(\pi_1, \pi_2, \dots, \pi_{S} \right)$; therefore $\sum_{s=1}^S \pi_s < 1$.  However, a statistically sufficient population is created, and then to preserve the probabilistic integrality (sum of the probabilities must equal 1), each probability is normalized using the softmax function for numerical stability due to the order of magnitude of the individual outcomes~\cite{bishop2006pattern}:
\begin{equation}
\pi_{s} = \dfrac{1}{K}\left(e^{\log \pi_{s} - \max\left(\log \boldsymbol{\pi}\right)}\right), \quad s=1,\dots,S,
\end{equation}
where is $K$ is a normalization constant given as: $K=\sum_{s=1}^Se^{\log\pi_{s} - \max\left(\log \boldsymbol{\pi}\right)}$.

\item Combining Probabilistic Scenarios:
Steps (i)-(iv) work for any probabilistic variable.  However, a probabilistic forecast could be given for any variable of interest; load $\ell$, wind $w$, and solar $\solar$, for example. Tools to determine operating reserves could use each of these on an itemized manner (Figs.~\ref{fig:exact_load}-\ref{fig:exact_solar}) to be later combined, or on an aggregated manner directly (i.e., for net demand, Fig.~\ref{fig:exact_net_load}).  If choosing the latter, then each element of the set would be given as:
\begin{equation}
\bd n_{i,j,k} = \boldsymbol{\ell}_i - \bd w_j - \boldsymbol{\solar}_k, \qquad \forall i,j,k,
\end{equation}
and there are as many scenarios in the set as combinations of $i,j$ and $k$.  Note that the sets of load, wind, and solar could have different cardinalities.  The probability of each of the elements in the aggregated set is then given as:
\begin{equation}
\pi^{n}_{i,j,k} = \pi^{{\ell}}_i \; \pi^{ w}_j \; \pi^{\solar}_k, \qquad \forall i,j,k.
\end{equation}
Considering all combinations of sets $i,j$ and $k$ ensures that $\sum \pi^{n}_{i,j,k} = 1$.
\end{enumerate}

\section{Dynamic Reserve Requirements}\label{sec:dynamic_reserve}
Dynamic Reserve requirements are those that vary over time based on specified conditions. These requirements can be used to allocate recourse to respond to uncertainty in a deterministic framework.  If determined correctly, they could bring forth the benefits of advanced scheduling models (e.g., stochastic scheduling) to their deterministic counterparts.  There are different approaches to determine operating reserves for flexibility (e.g.,~\cite{lew2013western,krad2015quantifying,ortega2020risk}), and a few tools on a greater readiness level (e.g., DynADOR, DRD, RESERVE). 

Exact reserve requirements are the differences between actual quantities and forecasted quantities.  These differences are denoted by $\epsilon$, and in the case of net demand ($n$) are given as:
\begin{equation}\label{eq:error}
\epsilon^{n}_h = n^{\rm A}_h - n^{\rm F}_h, \qquad h=1,2\ldots,H,
\end{equation}
where $H$ is the number of historical samples.

The exact reserve requirements can be itemized with respect to their direction (i.e., upward or downward) as: 
\begin{subequations}\label{eq:reserve_itemized}
\begin{alignat}{2}
\label{eq:reserve_up_itemized}
\boldsymbol{\epsilon}^{ n,\rm{up}} &= \left\{ \epsilon_h^{n} : \epsilon_h^{ n} > 0 \right\},\\
\label{eq:reserve_down_itemized}
\boldsymbol{\epsilon}^{ n,\rm{dn}} &= \left\{ -\epsilon_h^{ n} : \epsilon_h^{ n} < 0 \right\}.
\end{alignat}
\end{subequations}

Assuming an explanatory variable ${\nu}^{ n}$ is identified or chosen for the exact reserve requirements, (e.g., the magnitude of net demand $n^{\rm F}_h$, its rate of change $\Delta n^{\rm F}_h$, the hour of the day `$t$', or any other relevant quantity); the sets $\boldsymbol{\epsilon}^{n, \rm{up}}$ and $\boldsymbol{\epsilon}^{n, \rm{dn}}$ can be divided into $B$ subsets ordered as a function of ${\nu}^{n}$ as follows: 
\begin{subequations}\label{eq:reserve_subsets}
\begin{alignat}{2}
\boldsymbol{\epsilon}^{n, \rm{up}}_b &= \left\{ \epsilon_h^{ n} : l_{b-1} < \nu_i^{ n} \leq l_b,\; \epsilon_h^{ n} > 0 \right\},\\
\boldsymbol{\epsilon}^{n, \rm{dn}}_b &= \left\{ -\epsilon_h^{ n} : l_{b-1} < \nu_i^{ n} \leq l_b,\; \epsilon_h^{ n} < 0 \right\},
\end{alignat}
\end{subequations}
for $b=1,\dots,B$, where:
\begin{equation}\label{eq:bins}
l_b \! = \! \min\left( {\nu}^{ n} \right) + \tfrac{b}{B}  \left( \max\left( {\nu}^{ n} \right) - \min\left( {\nu}^{ n} \right) \right).
\end{equation}

Note that $B$ is a parameter that must be carefully chosen due to the bias-variance trade-off~\cite{bishop2006pattern}.  If $B$ is too small, the model will be under-fitted.  On the other hand, if $B$ is too large, the model will be over-fitted.

\begin{figure}[htb]
 \centering
 \begin{subfigure}[t]{1.72in}
     \centering
     \includegraphics[width=\textwidth]{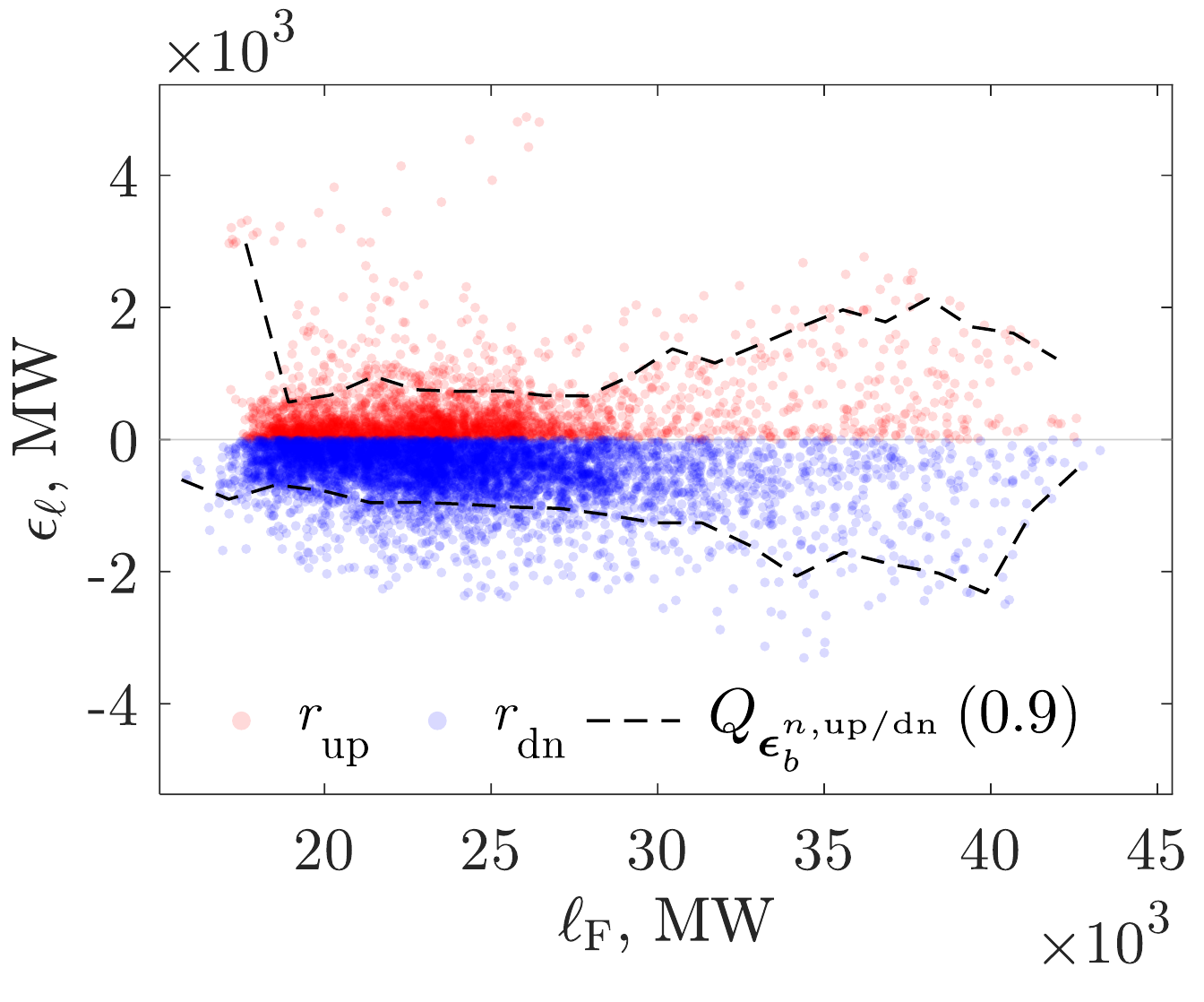}
     \caption{Historical reserve requirements for load as a function of load magnitude}
     \label{fig:exact_load}
 \end{subfigure}
 \hfill
 \begin{subfigure}[t]{1.72in}
     \centering
     \includegraphics[width=\textwidth]{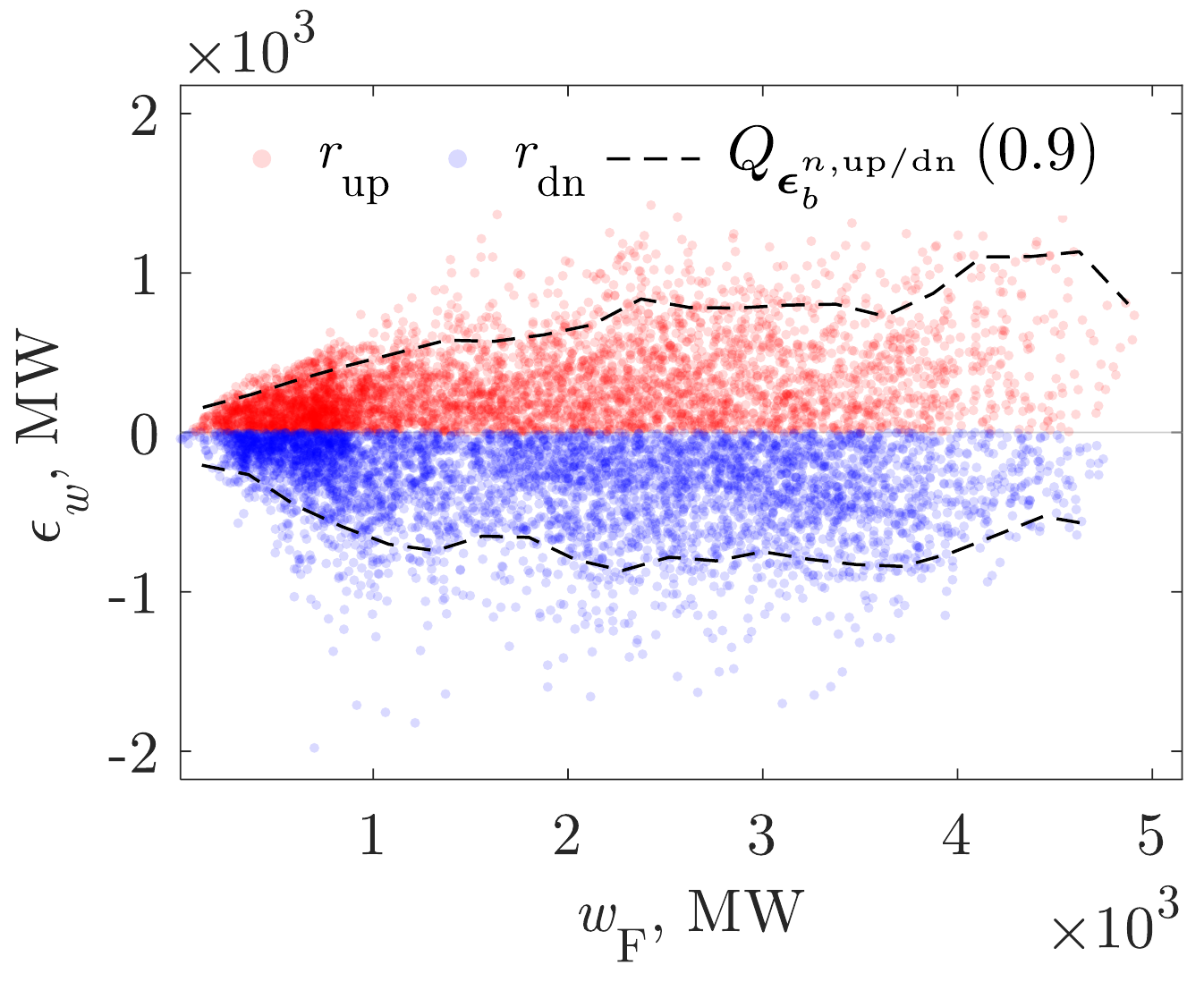}
     \caption{Historical reserve requirements for wind as a function of wind magnitude}
     \label{fig:exact_wind}
 \end{subfigure}
  \begin{subfigure}[t]{1.72in}
     \centering
     \includegraphics[width=\textwidth]{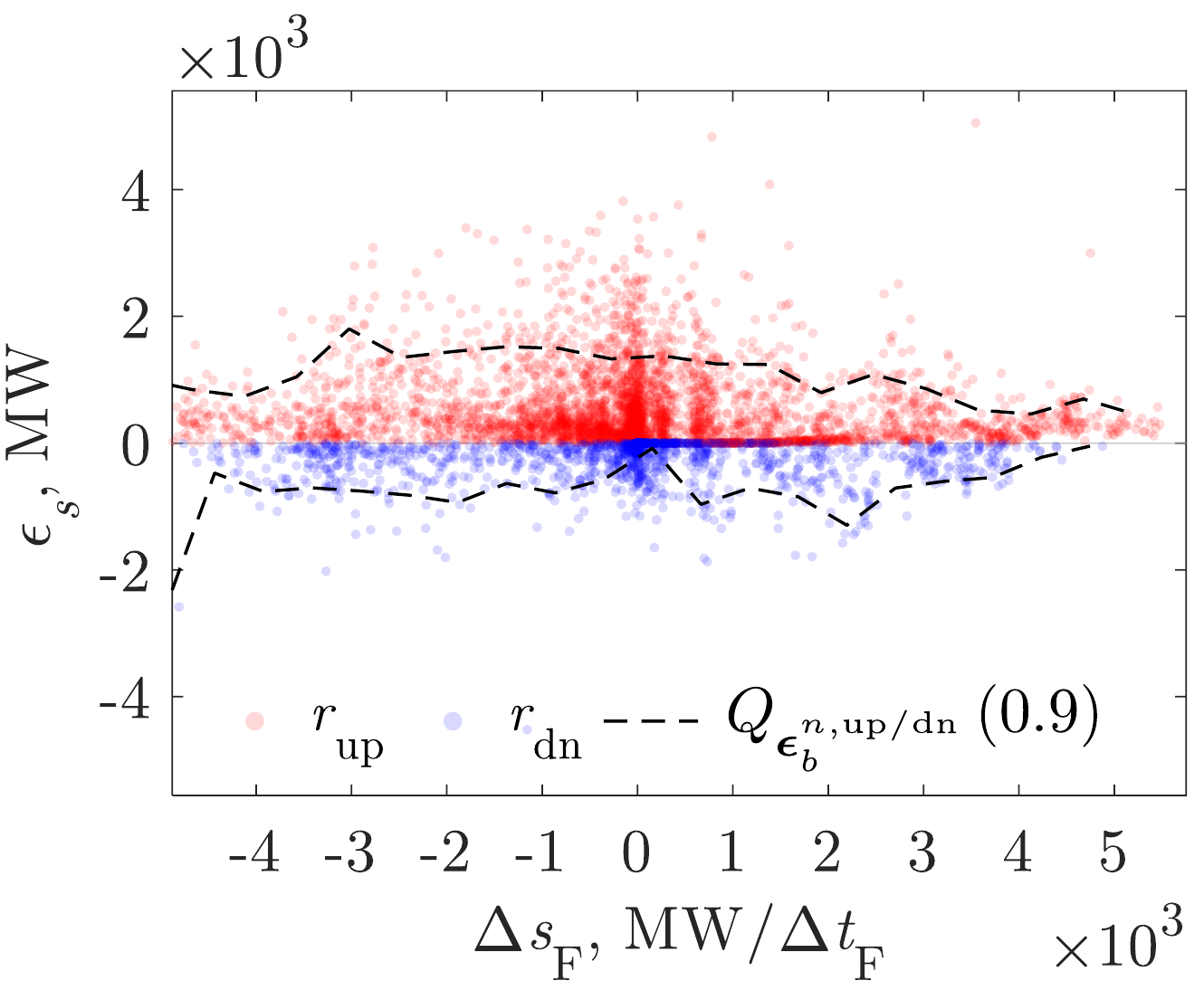}
     \caption{Historical reserve requirements for solar as a function of solar rate of change}
     \label{fig:exact_solar}
 \end{subfigure}
 \hfill
 \begin{subfigure}[t]{1.72in}
     \centering
     \includegraphics[width=\textwidth]{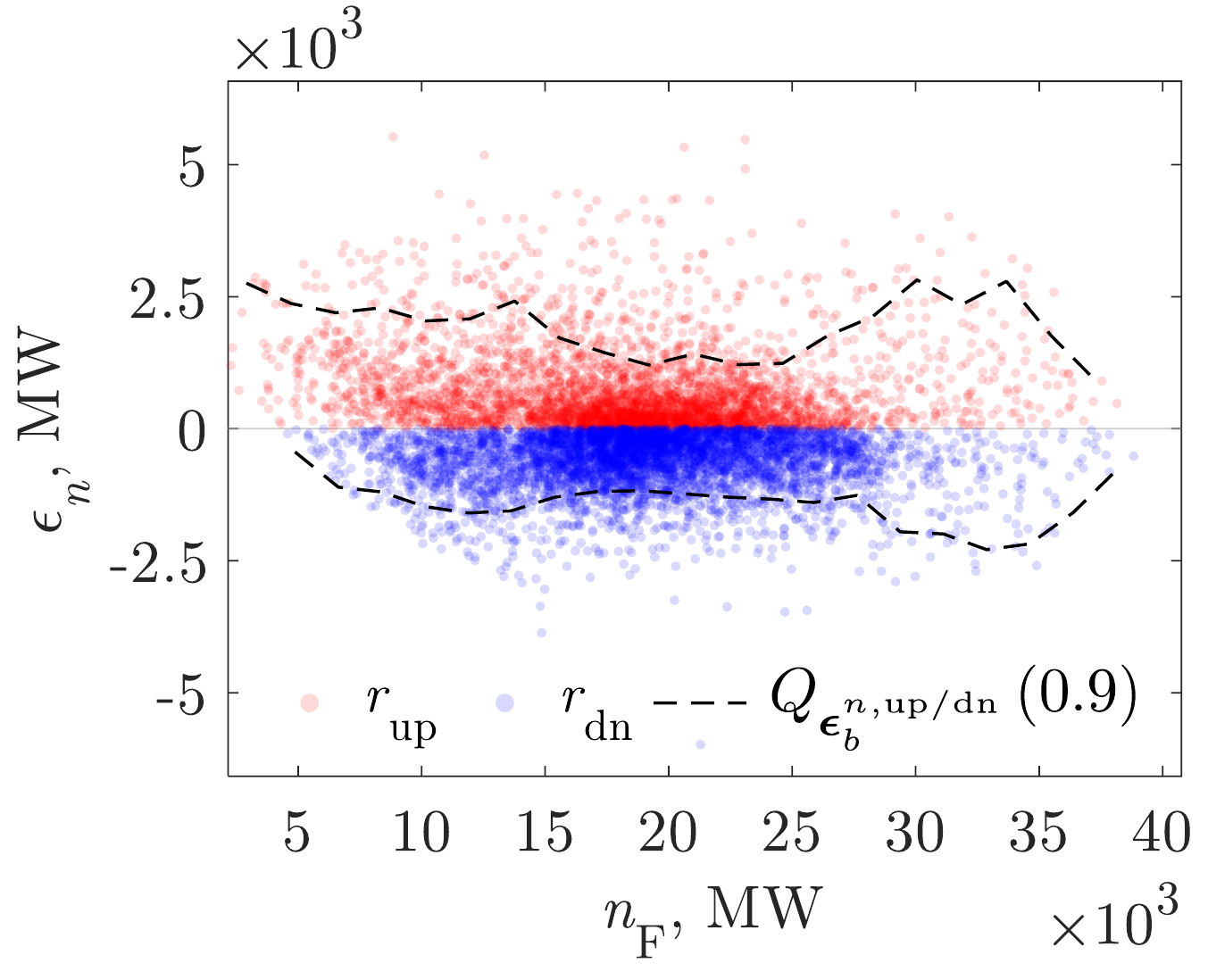}
     \caption{Historical reserve requirements for net demand as a function of net demand magnitude}
     \label{fig:exact_net_load}
 \end{subfigure}
    \caption{Historical exact reserve requirements by source for CAISO in $2019$.}
    \label{fig:exact_reserve}
\end{figure}

Using the exact historical reserve needs~\eqref{eq:reserve_subsets}, and given a desired confidence interval, $0 \leq {\rm CI} \leq 1$, the reserve requirements for an anticipated forecast can be determined using a quantile function $Q_{\boldsymbol{\epsilon}^{n, \rm{\text up/dn}}_b} \left({\rm CI}\right)$:
\begin{equation}\label{eq:reserve_requirements}
r_t^{n, \rm{up/dn}}\left( \nu_t^{n} \right)= 
\begin{cases}
Q_{\boldsymbol{\epsilon}^{n, \rm{\text up/dn}}_0}\left({\rm CI}\right) & \quad \nu_t^{n} \le l_0\\
Q_{\boldsymbol{\epsilon}^{ n, \rm{\text up/dn}}_b}\left({\rm CI}\right) & \quad l_{b-1} < \nu_t^{n} \leq l_b\\
Q_{\boldsymbol{\epsilon}^{n, \rm{\text up/dn}}_B}\left({\rm CI}\right) & \quad \nu_t^{n} > l_B ,
\end{cases}
\end{equation}
for $t=1,\dots,T$, where $T$ is the number of time intervals in the new forecast.
Fig.~\ref{fig:exact_reserve} illustrates the procedure to compute the exact historical reserve requirements for a confidence interval ${\rm CI}$, which is given by~\eqref{eq:error}-\eqref{eq:reserve_requirements}. In specific, Fig.~\ref{fig:exact_net_load} displays the exact historical reserves for net load. The red points correspond to the historical upward reserve requirements, given by~\eqref{eq:reserve_up_itemized}. Similarly, the blue points correspond to the historical downward reserve requirements, given by~\eqref{eq:reserve_down_itemized}. Then, these historical reserve requirements are itemized and computed according to~\eqref{eq:reserve_subsets}, \eqref{eq:bins}, and the quantile function $Q_{\boldsymbol{\epsilon}^{n, \rm{\text up/dn}}_b} \left({\rm CI}\right)$ with $B=20$ and ${\rm CI}=0.9$. As result, the reserve requirements for a given level of forecasted net load are obtained, and they are delineated by the dashed black lines in Fig.~\ref{fig:exact_net_load}.
%
There are other alternatives to compute these reserve requirements, such as gradient boosting quantile regression~\cite{yuan2015random} or non-parametric quantile regression~\cite{takeuchi2006nonparametric}. But, the simpler model given by~\eqref{eq:reserve_subsets}-\eqref{eq:reserve_requirements} is chosen because the objective of this work is to show how to present methods to capitalize on the uncertainty information from probabilistic forecasts for reserve determination.

Equations~\eqref{eq:error}-\eqref{eq:reserve_requirements} may also be used to quantify the reserve requirements by source, that is, load $\ell$, wind $w$, or solar $\solar$~\cite{lew2013western,krad2015quantifying}.
Then, with these reserve requirements by source and assuming that there is no correlation between them, the total reserve requirements are expressed as follows~\cite{athanasios2002probability}:
\begin{subequations}\label{eq:reserve_sum}
\begin{alignat}{2}
r_t^{\rm{up}} &= \left[
\left(r_t^{{\ell}, \rm{up}}\right)^2 + \left(r_t^{ w, \rm{up}}\right)^2 + \left(r_t^{ \solar, \rm{up}}\right)^2
\right]^\frac{1}{2},\\
r_t^{\rm{dn}} &= \left[
\left(r_t^{{\ell}, \rm{dn}}\right)^2 + \left(r_t^{ w, \rm{dn}}\right)^2 + \left(r_t^{ \solar, \rm{dn}}\right)^2\right]^\frac{1}{2}.
\end{alignat}
\end{subequations}

Note that according to the net load definition ($n_t = \ell_t -  w_t - \solar_t$), a positive value in~\eqref{eq:error} for load corresponds to needs for up reserves, and a negative one corresponds to needs for down reserves.
On the contrary, for wind and solar generation, a positive value in~\eqref{eq:error} corresponds to needs for down reserves. A negative one corresponds to needs for up reserves, as the net load expression implies. 
\section{Probabilistic Dynamic Reserve Requirements}\label{sec:probabilistic_reserve}
Using the dynamic reserve requirement method presented in the previous section, there are different ways in which probabilistic forecasts could be used to determine probabilistic dynamic reserve requirements.  In this section, these methods are categorized into three groups as follows:
\begin{enumerate}[i)]
\item {Recursive methods}: These methods are based on historical information on the characteristics of the deviations for different forecasts and operating conditions.  In these models, the probabilistic information (from the anticipated forecast) is compared against the historical deviations to predict the reserve requirements needed to accommodate future deviations.  Most of the reserve determination methods are based on recursive approaches, e.g., ~\cite{bruninx2014statistical, de2019dynamic, cauwet2019static, ortega2020risk}.
These methods perform well when the system undergoes deviations that share similar characteristics with previously observed data (e.g., morning and evening ramps).  These methods do not respond well to previously ``unseen'' events (e.g., extremely cloudy days with low temperature).
\item {Anticipative methods:} These methods only use the probabilistic forecasts of anticipated conditions to determine the reserve requirements.  That is, all the reserve requirements are based on the information that the probabilistic forecast is offering, and this information is not compared against any historical equivalent conditions.  This could be a drawback if the system undergoes similar and repeated deviations that are not captured in the probabilistic forecast.  However, these methods could excel at capturing `unique' events on a day-to-day basis (e.g., weather-driven events).  It is expected that tools to determine reserves for uncertainty will rely more heavily on probabilistic forecasts as these tools mature, and ISO, utilities, and industry, in general, embrace their use and application. 
\item {Hybrid methods:} These methods rely on both historical information as well as probabilistic information of anticipated conditions to determine the reserve requirements.  These methods combine the benefits of the recursive and the anticipative methods by procuring reserves that protect against historical conditions that are somewhat periodic, as well as those conditions that are specific to any given day.   
\end{enumerate}
The following subsections present two recursive approaches, two anticipative approaches, and a hybrid approach.
\subsubsection{All Scenarios [Recursive]} This is the most intuitive form of using the probabilistic forecast, which consists in using all the information obtained from the conversion between probabilistic forecast and probabilistic scenarios, Fig.~\ref{fig:probabilistic_scenarios}.  This entails passing each of the scenarios through the dynamic reserve determination process (section~\ref{sec:dynamic_reserve}) and obtaining the upward and downward reserve requirements for such scenario $r_s^{\rm up}$ and $r_s^{\rm dn}$.  The expected requirements for the complete set of scenarios are:
\begin{subequations}\label{eq:reserve_scenarios}
\begin{equation}\label{eq:rup_prob}
\mathbf{r}^{\rm up} = \sum_{s=1}^S \pi_s \mathbf{r}_s^{\rm up},
\end{equation}
\begin{equation}\label{eq:rdn_prob}
\mathbf{r}^{\rm dn} = \sum_{s=1}^S \pi_s \mathbf{r}_s^{\rm dn}. 
\end{equation}
\end{subequations}

\subsubsection{Extreme Scenarios [Recursive]}
From an obtained set of scenarios, it is possible to obtain the extreme realizations with respect to a given feature or explanatory variable (e.g., production, rate of change, etc.).  This is done by assigning a score to each scenario $\bd s_i$ and then aggregating the values from their explanatory variable $\boldsymbol{\nu}^{\bd s_s}$:
\begin{equation}\label{eq:score_scenarios}
\gamma_s = \sum_{t=1}^{T} \nu^{\bd s_s}_{t},
\end{equation}
where $\gamma_i$ is the sum of aggregated values over the horizon associated with scenario $\bd s_s$.
Then, the set of all the aggregated values or scores is defined as $\Gamma = \left\{ \gamma_1, \gamma_2, \dots, \gamma_{S} \right\}$. 
The $n$ most extreme up scenarios are the scenarios associated with the $d$ largest values in $\Gamma$, Fig.~\ref{fig:extreme_up}.
Similarly, the $d$ most extreme down scenarios are those associated with the $d$ lowest values in $\Gamma$, Fig.~\ref{fig:extreme_down}.  Note that in Fig.~\ref{fig:extreme_scenarios} the scenarios are chosen for solar generation, resulting in low production scenarios associated with upward reserves and high production scenarios for downward reserves.  Given the worst upward and downward subsets of scenarios (i.e., $\bd{S}^{\rm up}$ and $\bd{S}^{\rm dn}$), it is then possible to compute the upward reserve requirements using equation~\eqref{eq:rup_prob} for the worst upward scenarios $\mathbf{S}^{\rm up}$, and the downward requirements using equation~\eqref{eq:rdn_prob} for the worst downward scenarios $\mathbf{S}^{\rm dn}$.

\subsubsection{Bounds of Extreme Scenarios [Anticipative]}
A variant that relies on extreme scenarios consists in determining the expected worst upward and downward scenarios directly from sets $\mathbf{S}^{\rm up}$ and $\mathbf{S}^{\rm dn}$, respectively. This is done by computing the expected scenario from each subset as follows:
\begin{subequations}\label{eq:E_bounds}
\begin{equation}\label{eq:sup_bounds}
\mathbf{s}^{\rm up} = \dfrac{1}{\sum_{s=1}^{S^{\rm up}} \pi_s} \sum_{s=1}^{S^{\rm up}} \pi_s \mathbf{s}_s^{\rm up},
\end{equation}
\begin{equation}\label{eq:sdn_bounds}
\mathbf{s}^{\rm dn} = \dfrac{1}{\sum_{s=1}^{S^{\rm dn}} \pi_s} \sum_{s=1}^{S^{\rm dn}} \pi_s \mathbf{s}_s^{\rm dn}. 
\end{equation}
\end{subequations}

Then, the upward and downward reserve requirements are given as the difference of the expected up and down scenarios with respect to the central forecast:
\begin{subequations}\label{eq:reserve_bounds}
\begin{equation}\label{eq:rup_bounds}
\mathbf{r}^{\rm up} = \mathbf{s}^{\rm cf} - \mathbf{s}^{\rm up},
\end{equation}
\begin{equation}\label{eq:rdn_bounds}
\mathbf{r}^{\rm dn} = \mathbf{s}^{\rm dn} - \mathbf{s}^{\rm cf},
\end{equation}
\end{subequations}
where $\mathbf{s}^{\rm cf}$ is the central forecast. Fig.~\ref{fig:reserve_bounds} depicts the concept for a solar probabilistic forecast.

\subsubsection{Prediction Interval (PI) [Anticipative]}  
\begin{figure}[htb]
 \centering
 \begin{subfigure}[t]{1.72in}
     \centering
     \includegraphics[width=\textwidth]{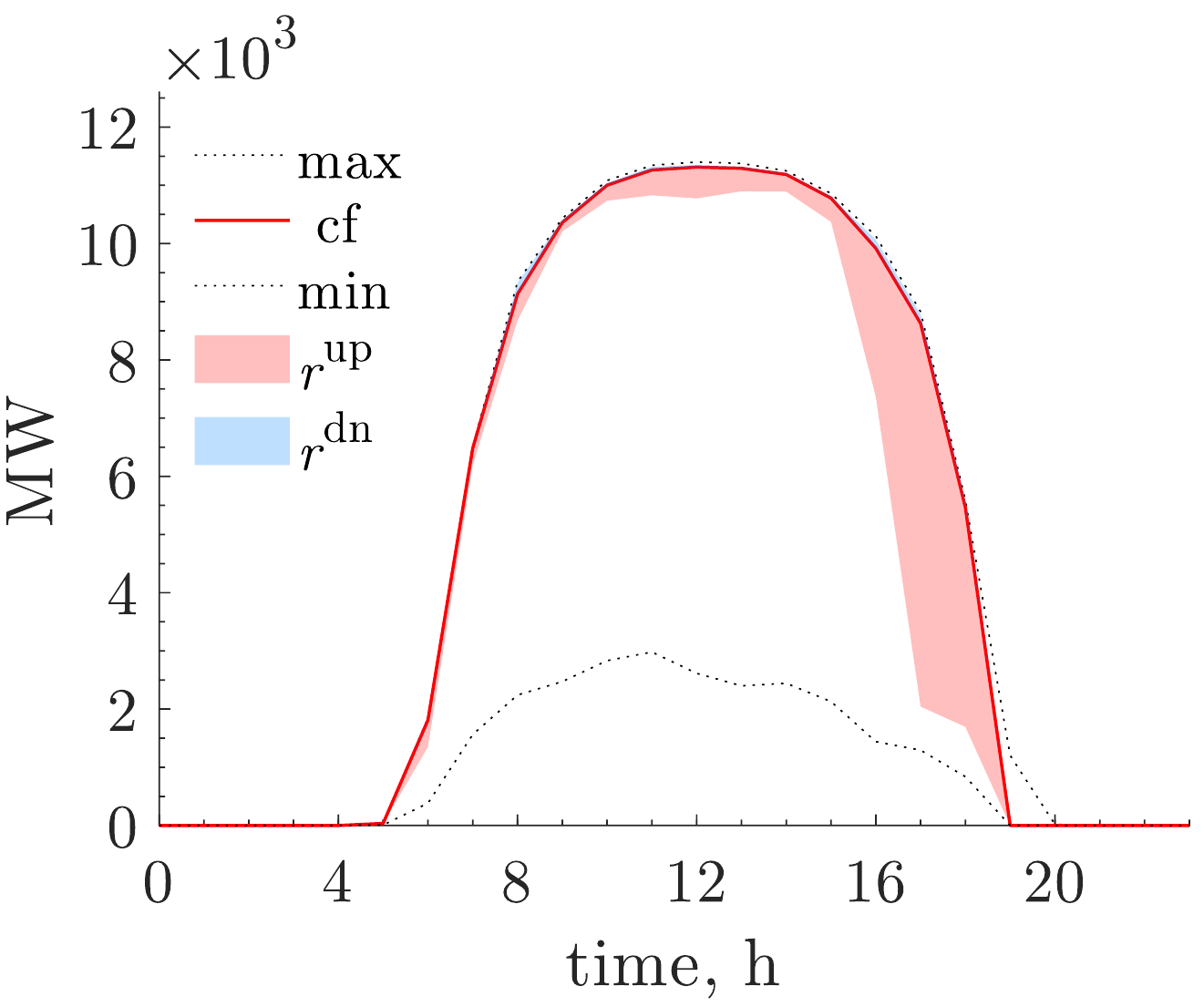}
     \caption{Bounds of extreme scenarios}
     \label{fig:reserve_bounds}
 \end{subfigure}
 \hfill
 \begin{subfigure}[t]{1.72in}
     \centering
     \includegraphics[width=\textwidth]{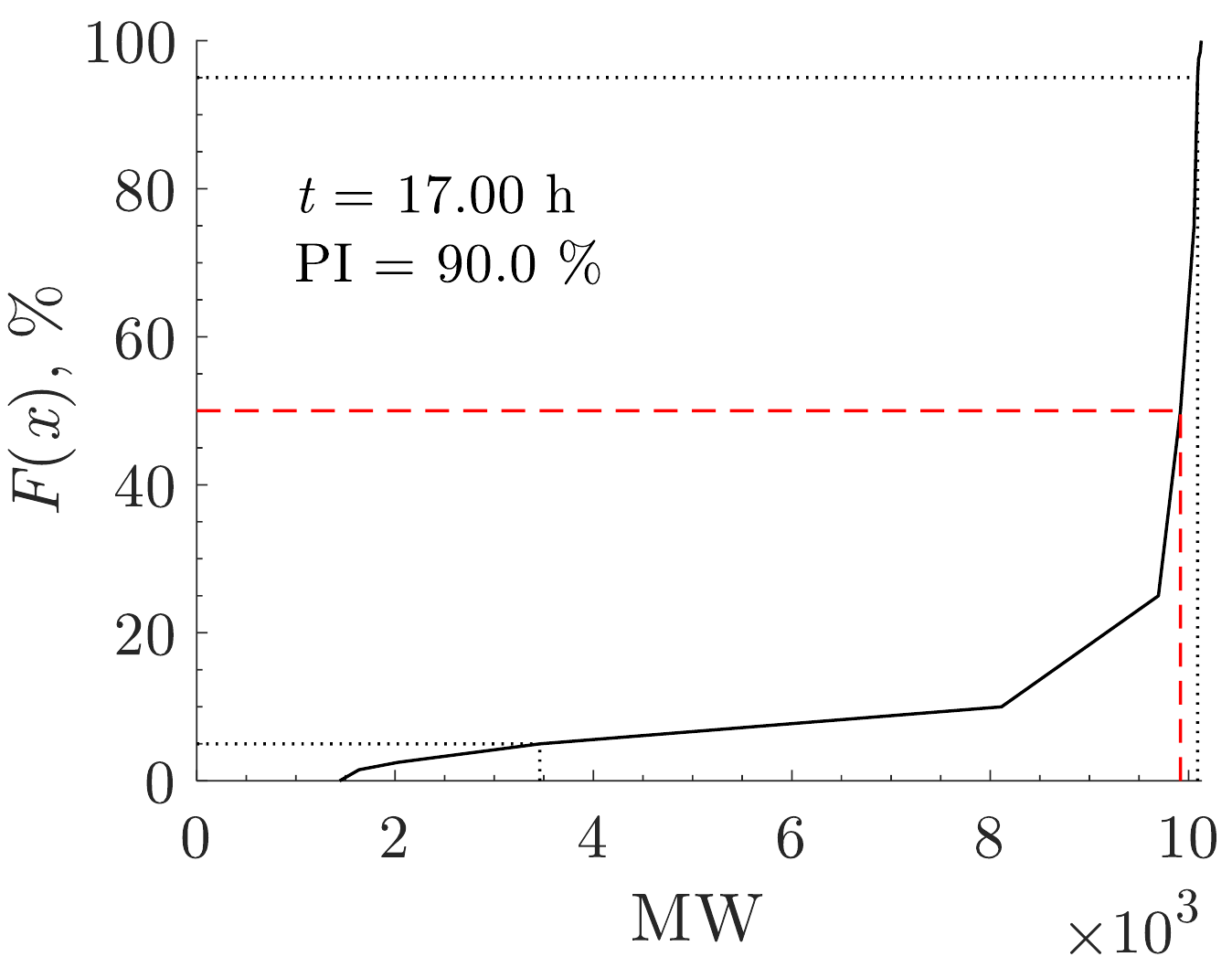}
     \caption{Prediction interval}
     \label{fig:reserves_PI}
 \end{subfigure}
    \caption{Reserve assessment with anticipative methods.}
    \label{fig:reserves_bounds_PI}
\end{figure}

This method consists in using the CDF ($F(\cdot)$) of each time sample of the prediction interval to enforce a desired PI.  The PI represents a percentage of the uncertainty to protect the against.  The chosen ${\rm PI}$ result in a probabilistic limit of $P_{\rm limit} = \left( \tfrac{1-{\rm PI}}{2} \right) \, 100\%$.  Then the reserve requirements are, Fig.~\ref{fig:reserves_PI}:
\begin{subequations}\label{eq:reserve_PI}
\begin{alignat}{2}
\label{eq:rup_PI}
r^{\rm up}_t &= F^{-1}_t(1-P_{\rm limit}), \quad t=1,\dots,T,\\
\label{eq:rdn_PI}
r^{\rm dn}_t &= -F^{-1}_t(P_{\rm limit}), \quad t=1,\dots,T.
\end{alignat}
\end{subequations}
%

\subsubsection{Hybrid Method [Both]} This method combines any of the methods described above or any other alternative available to the user.  It could combine `$M$' methods ($m_M$) by taking the most conservative reserve requirement value among the $M$ methods at each time interval.  In this case, the reserve requirements are given as the maximum reserve requirement value:
\begin{subequations}\label{reserve_hybrid}
\begin{equation}\label{eq:rup_hybrid}
r^{\rm up}_t = {\rm max}\left\{r^{\rm up}_{t,m_1}, \dots, r^{\rm up}_{t,m_M} \right\}, \quad t=1,\dots,T,
\end{equation}
\begin{equation}\label{eq:rdn_hybrid}
r^{\rm dn}_t = {\rm max}\left\{r^{\rm dn}_{t,m_1}, \dots, r^{\rm dn}_{t,m_M} \right\}, \quad t=1,\dots,T.
\end{equation}
\end{subequations}
%

\section{Test Results}\label{sec:results}
The proposed approaches to determine probabilistic dynamic reserve requirements have been developed as part of the DOE-funded project ``Operational Probabilistic Tools for Solar Uncertainty (OPTSUN)''~\cite{tuohy2019operational}.
The California ISO provided the datasets for load, wind, and solar. In specific, CAISO provided the Day-Ahead Market (DAM) forecasted data with a $60$-minute resolution and Fifteen Minute Market (FFM) data with a $15$-minute resolution for load, solar, and wind generation for $2019$,~\cite{CAISOtoday2021}.
UL, who is also part of the OPTSUN team, provided the corresponding probabilistic forecasts, as well as forecasts for $2020$.

To test the proposed recursive, anticipative, and hybrid methods, July $1^{\rm st}$, $2020$ is selected as a study case, being a day with mixed weather conditions.  These mixed weather conditions will show the differences between recursive and anticipative methods.  Only the probabilistic forecast for solar generation is used in order to isolate the effects on the reserve requirements and to demonstrate the accuracy of the scenario generation method, Fig.~\ref{fig:probabilistic_forecast}.  Deterministic forecasts are used for load and wind generation. For each of the experiments, $1000$ scenarios were generated with $\theta = 0.92$ and $\omega = 0.42$, which were obtained recursively from the $2019$ dataset of solar generation.

\subsection{Statistical Validation of Probabilistic Scenarios}
\begin{figure}[htb]
 \centering
 \begin{subfigure}[t]{1.72in}
     \centering
     \includegraphics[width=\textwidth]{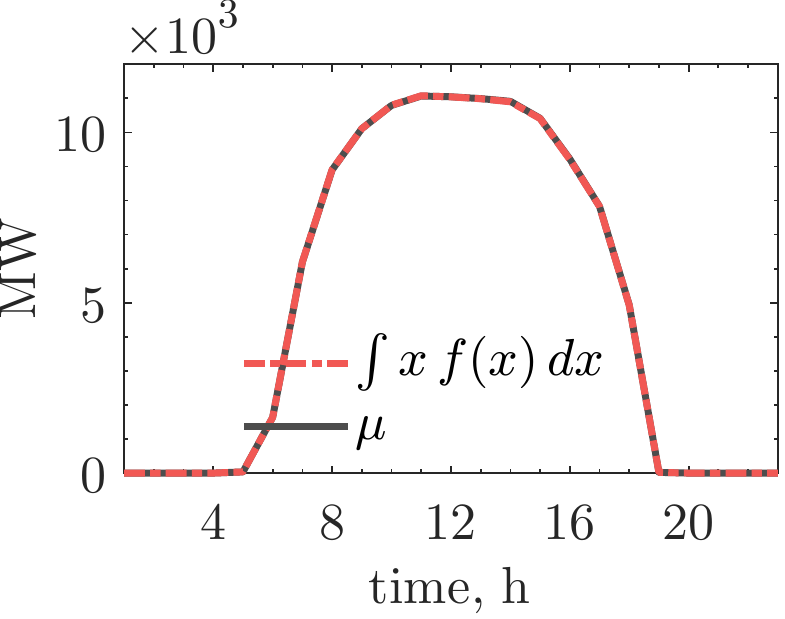}
     \caption{Mean}
     \label{fig:first_moment}
 \end{subfigure}
 \hfill
 \begin{subfigure}[t]{1.72in}
     \centering
     \includegraphics[width=\textwidth]{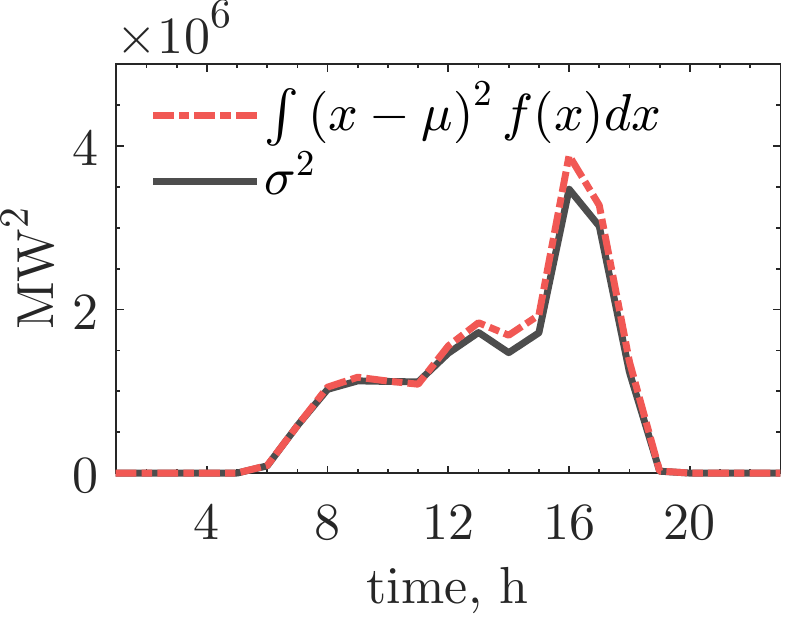}
     \caption{Variance}
     \label{fig:second_moment}
 \end{subfigure}
  \begin{subfigure}[t]{1.65in}
     \centering
     \includegraphics[width=\textwidth]{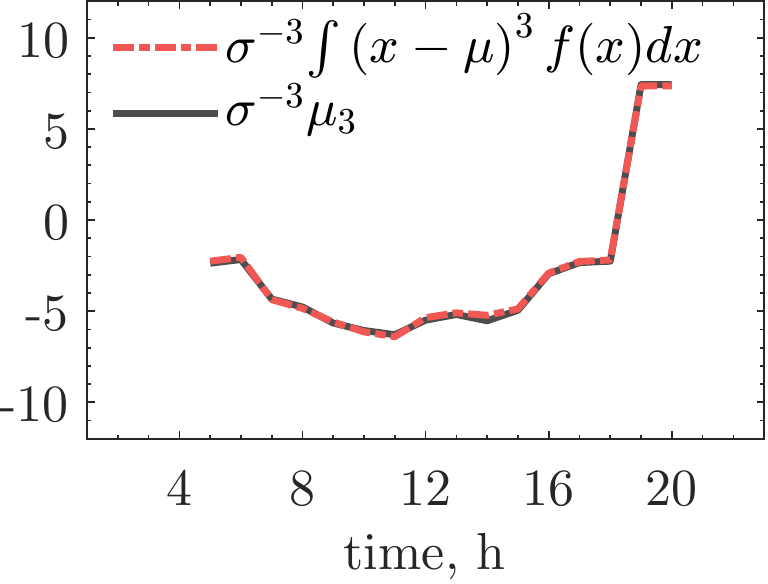}
     \caption{Skewness}
     \label{fig:third_moment}
 \end{subfigure}
\hspace{2.5mm}
 \begin{subfigure}[t]{1.65in}
     \centering
     \includegraphics[width=\textwidth]{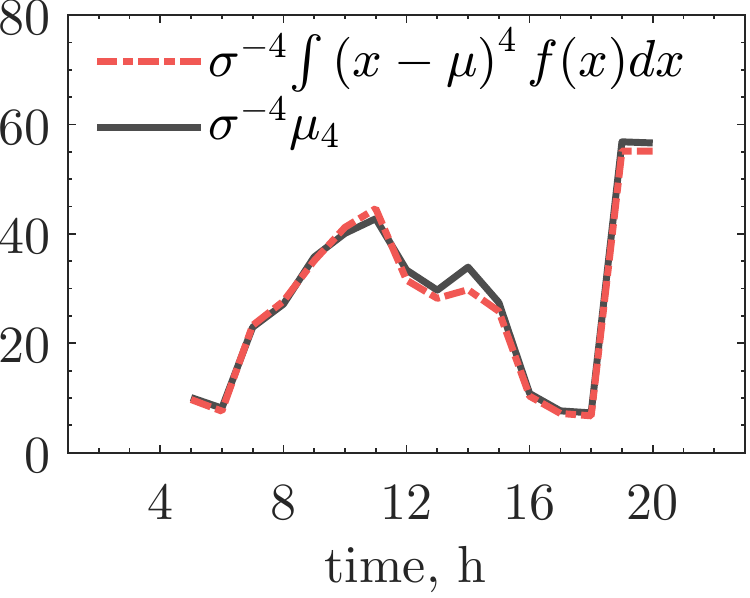}
     \caption{Excess kurtosis}
     \label{fig:fourth_moment}
 \end{subfigure}
    \caption{Mathematical moments of the probabilistic forecast and set of $1,000$ generated scenarios.}
    \label{fig:moment_scenarios}
\end{figure}

Looking at Fig. \ref{fig:probabilistic_forecast} and Fig. \ref{fig:probabilistic_scenarios}, a natural question is: ``how does the statistical information contained in the probabilistic forecast compare against the probabilistic scenarios?''. 
Since probabilistic forecasts are given as a set of non-parametric PDFs for each time interval, in the form of probability thresholds, it is then possible to compute the mathematical moments of these random variables at each period~\cite{athanasios2002probability}.  Similarly, given a set of probabilistic scenarios, it is possible to calculate the mathematical moments~\cite{joanes1998comparing} at each period. 
Fig.~\ref{fig:moment_scenarios} shows the mean, variance, skewness, and excess kurtosis for the probabilistic forecasts and a set of $1000$ probabilistic scenarios.  Note that as more scenarios are generated, the moments will converge to the moments of the probabilistic forecast with greater precision.  The following observations are made:
\begin{itemize}
\item[1)] Mean \& central forecast ($\mu$): In Fig.~\ref{fig:first_moment} it can be seen that the mean of the probabilistic forecast and the one obtained with the scenarios are virtually identical since the aggregated error between them is less than 0.1\% ($\%{\rm NRMSE} = 0.071\%$).

\item[2)] Variance ($\sigma^2$): is a measure of the dispersion from the mean, and in Fig.~\ref{fig:second_moment} it can be seen that these not only follow the same trends but also the difference between them is relatively small ($\%{\rm NRMSE} = 1.465\%$). In general, the variance is high between $15$ and $18$ h because the probabilistic forecast (Fig.~\ref{fig:probabilistic_forecast}) features thresholds of high probability dispersion from the central forecast.

\item[3)] Skewness ($\frac{\mu_3}{\sigma^3}$): measures the distributions' asymmetry. 
In Fig.~\ref{fig:third_moment}) it can be seen that there are only small differences between the two ($\%{\rm NRMSE} = 0.743 \%$).  Also, between $5$ and $18$ h, the skewness is negative since the distribution is highly skewed below the central forecast.  Note that the skewness is not defined during night hours because there is zero solar production expected during these periods.  

\item[4)] Excess kurtosis ($\frac{\mu_4}{\sigma^4}$): is a measure of the ``tailedness'' of the probability distribution.  
In Fig.~\ref{fig:fourth_moment} it can be seen that, again, the difference between the computed values is relatively small ($\%{\rm NRMSE} = 2.517\%$). In terms of shape, the largest value of the kurtosis is at $19$ h since this is the period in which a larger amount of outliers are expected.
\end{itemize}
%

\subsection{Reserve Requirements}
\begin{figure}[htb]
 \centering
 \begin{subfigure}[t]{1.72in}
     \centering
     \includegraphics[width=\textwidth]{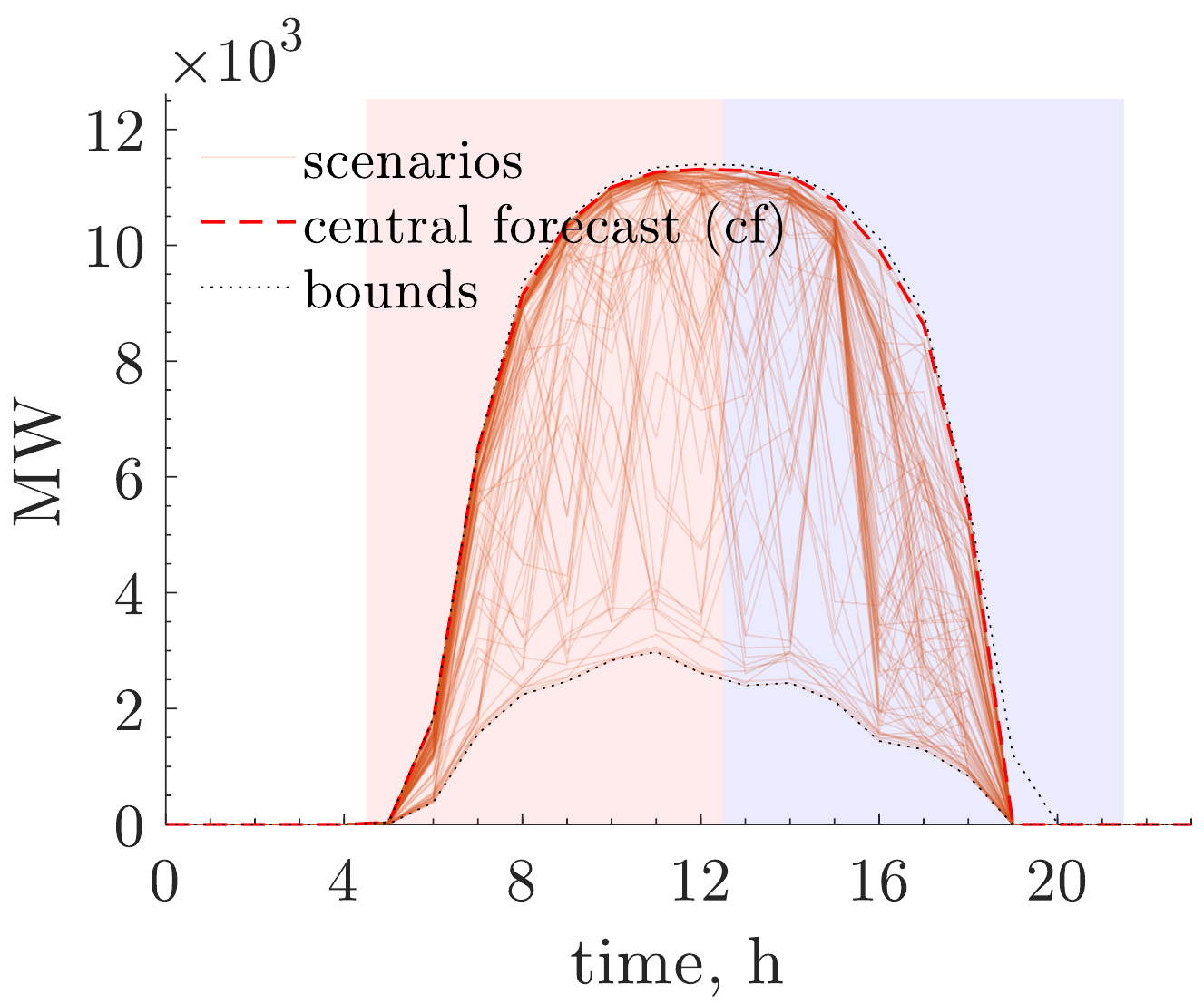}
     \caption{Upward extreme scenarios}
     \label{fig:extreme_up}
 \end{subfigure}
 \hfill
 \begin{subfigure}[t]{1.72in}
     \centering
     \includegraphics[width=\textwidth]{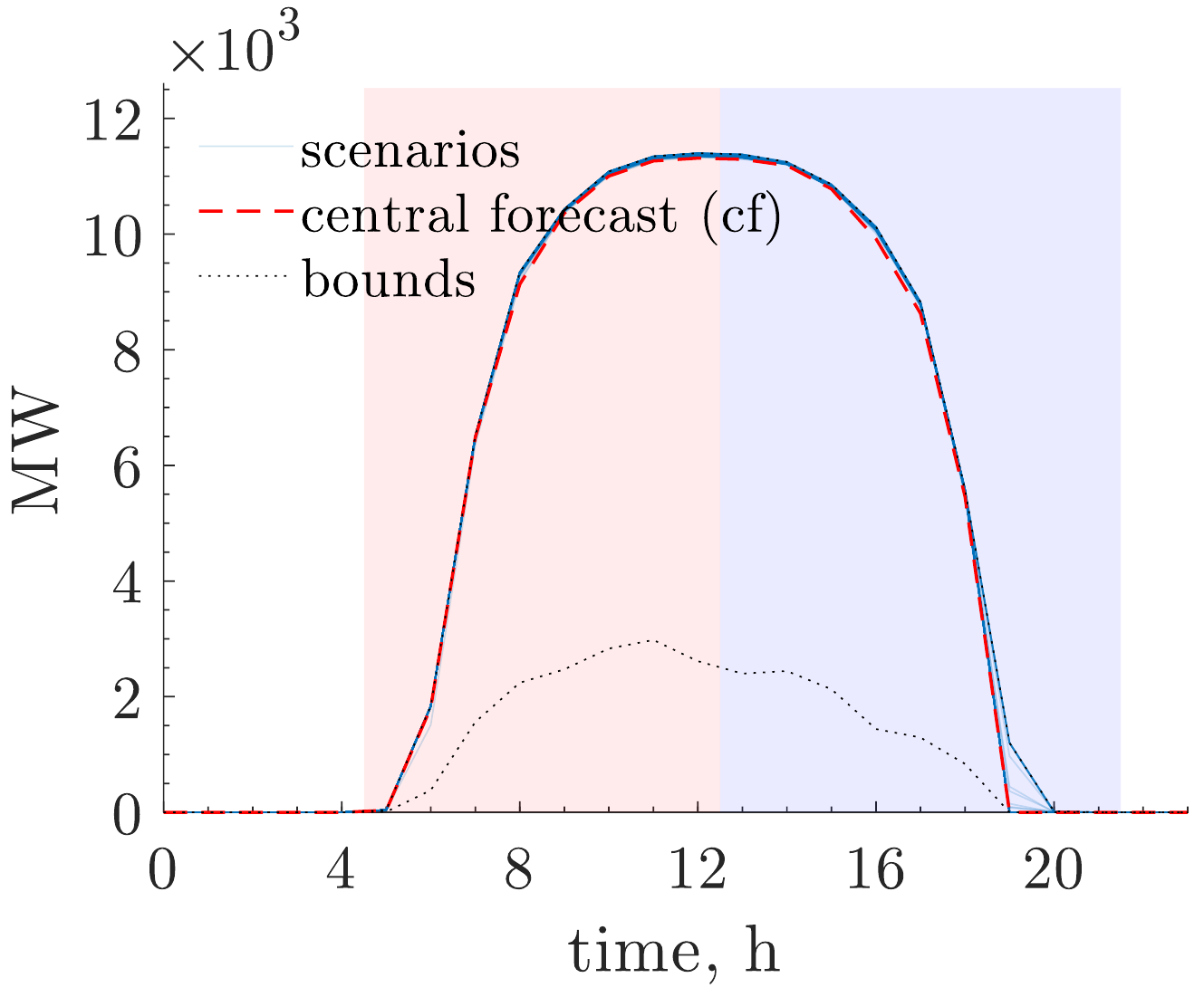}
     \caption{Downward extreme scenarios}
     \label{fig:extreme_down}
 \end{subfigure}
    \caption{Visualization of extreme scenarios.}
    \label{fig:extreme_scenarios}
\end{figure}

\begin{figure}[htb]
 \centering
 \begin{subfigure}[t]{1.72in}
     \centering
     \includegraphics[width=\textwidth]{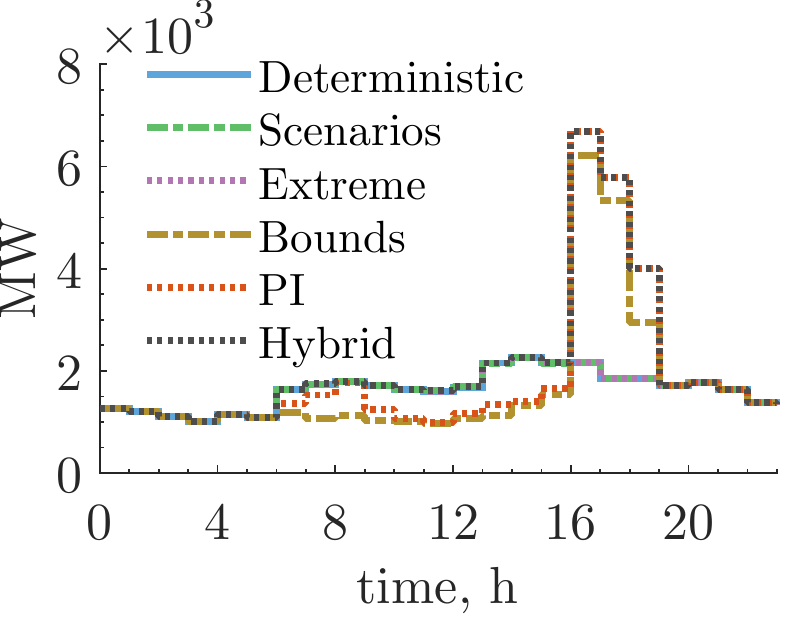}
     \caption{Reserve up.}
     \label{fig:reserve_up}
 \end{subfigure}
 \hfill
 \begin{subfigure}[t]{1.72in}
     \centering
     \includegraphics[width=\textwidth]{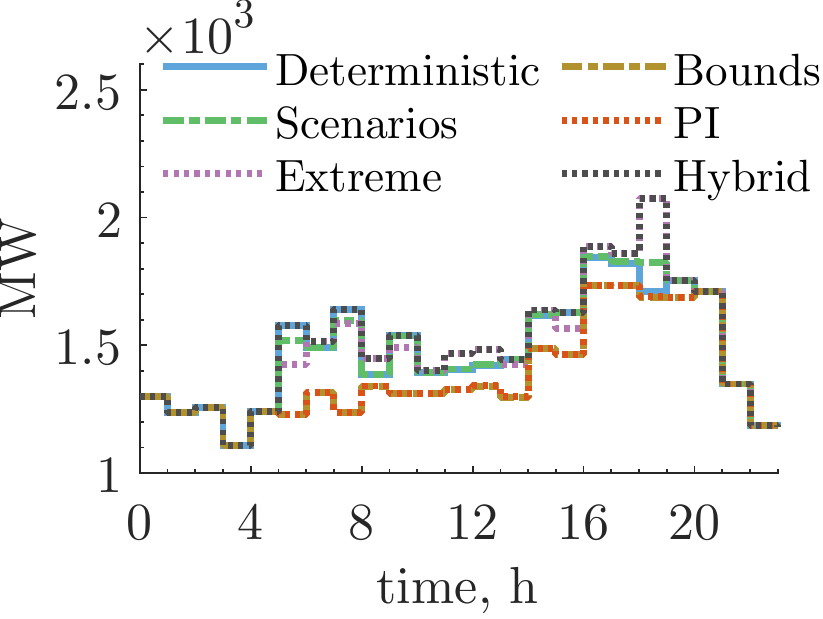}
     \caption{Reserve down.}
     \label{fig:reserve_down}
 \end{subfigure}
    \caption{Reserve assessment with different methods.}
    \label{fig:reserves}
\end{figure}

In this subsection, the up and down reserve requirements are presented using all the proposed approaches, Fig.~\ref{fig:reserves}.
The base case consists in assessing the reserve requirements as in~\eqref{eq:error}-\eqref{eq:reserve_sum}.  In that case, the explanatory variables for load and wind are the magnitude of the forecast ($ \ell^{\rm F}$ and $ w^{\rm F}$); and for the solar generation, the rate of change of the forecast ($\Delta \solar^{\rm F}$).  Also, a confidence interval (CI) of $90\%$ is enforced, and $20$ bins are used to partition the historical exact reserve requirements; The value of $B=20$ was found to have a good bias-variance trade-off, as shown in Fig.~\ref{fig:exact_reserve}.
\begin{itemize}
\item \textit{Deterministic}.
In this case only the central forecast for solar is used, (red line in Fig.~\ref{fig:probabilistic_forecast}).  The reserve requirements for each stochastic variable (i.e., $\ell$, $w$, and $\solar$) are determined using eq.~\eqref{eq:reserve_requirements}, the total reserve requirements for the system are computed using eq.~\eqref{eq:reserve_sum}, and shown with a blue solid line in Fig.~\ref{fig:reserves}.
\item \textit{All Scenarios}.  
For this case a set of $1000$ solar scenarios is obtained as discussed in Section~\ref{sec:PF_scenarios}, Fig.~\ref{fig:probabilistic_scenarios}. The reserve requirements are computed with eq.~\eqref{eq:reserve_scenarios}, and the results are shown in Fig.~\ref{fig:reserves} with a green dash-dot line.
These results are similar to the ones from the deterministic method.  This is expected since the average reserve requirements should converge to those obtained with the central values.
\item \textit{Extreme Scenarios}
In this case, the scenarios are weighted as show in equation~\eqref{eq:score_scenarios}, then the $10\%$ of the worst upward and downward scenarios are selected, $\bd{S}^{\rm up}$ and $\bd{S}^{\rm dn}$, Fig.~\ref{fig:extreme_scenarios}.  Fig.~\ref{fig:extreme_up} shows that the extreme upward scenarios are below the central forecast, and they have a significant dispersion within the probabilistic forecast.  On the other hand, the extreme downward scenarios shown in Fig.~\ref{fig:extreme_down} are above and virtually on top of the central forecast due to the extremely low dispersion of the probabilistic forecast in this direction.  The results for the upward and downward reserves are shown in Fig.~\ref{fig:reserves} with a purple dotted line.
\end{itemize}

These three methods rely upon historical data to determine the reserve requirements (i.e., recursive methods). 
Because of this reason, the resulting requirements follow the same trends, as shown in Fig.~\ref{fig:reserves}.  However, between $16$--$18$ h, the method based on extreme scenarios allocates a larger amount of reserve because the extreme scenarios are quite different from the central forecast for those intervals, Fig.~\ref{fig:reserve_down}. 
%

\begin{itemize}
\item \textit{Bounds of Extreme Scenarios}.
Given the set of extreme scenarios $\bd{S}^{\rm up}$ and $\bd{S}^{\rm dn}$, their expected value is given~\eqref{eq:E_bounds}, and with this it is then possible to compute the solar reserve requirements as~\eqref{eq:reserve_bounds}, and the combine using~\eqref{eq:reserve_sum}, see Fig.~\ref{fig:reserves} brown dash-dot line
It can be seen that this approach is more sensitive to the shape of the probabilistic forecast.
\item \textit{Prediction Interval}.
In this case the prediction interval is $\rm{PI} = 0.9$. 
The solar reserve requirements are determined using~\eqref{eq:reserve_PI}, and the total system reserves are computed with~\eqref{eq:reserve_sum}.  Fig.~\ref{fig:reserves} shows the results for this case using an orange dotted line.
\end{itemize}

Both of these approaches are anticipative as they rely exclusively on the probabilistic forecast.
 For the downward reserve requirements in Fig.~\ref{fig:reserve_down}, these anticipative methods allocate low amounts of reserve than the recursive methods. This is because according to the probabilistic forecast, the solar generation above the central forecast is virtually unlikely and of low probability, Fig.~\ref{fig:probabilistic_forecast}.

For the upward reserve requirements, Fig.~\ref{fig:reserve_up}, with the exception of the interval $16$--$18$ h, the reserve requirements are lower or equal to the recursive methods.  This is because, historically, there have been larger deviations due to solar generation than those captured for this particular day with the probabilistic forecast.  These requirements also produce a large peak 
between $16$--$18$ h, because the probabilistic forecast indicates a large probability of reducing the solar power production, Fig.~\ref{fig:probabilistic_forecast}. 

This example extols the benefits of relying on certain probabilistic information, as this would hedge the system against events with a large probability of materialization on a particular day, as opposed to taking probabilistic quantiles over long historical horizons. 

\begin{itemize}
\item \textit{Hybrid Method}.
As explained, this method consists in taking the largest values for any combination of methods.  In this case, all the presented methods are merged, and the upward and downward reserves are shown in Fig.~\ref{fig:reserves} on a grey dotted line.
By combining recursive and anticipative approaches, the resulting reserves would hedge the system against highly expected deviation based on historical information, as well as highly expected conditions based on weather patterns for the day/period under study. 
\end{itemize}

Comparing all the methods shown in Fig.~\ref{fig:reserve_up}, it can be seen that between $6$--$15$ h, the anticipative methods allocate less reserve than the recursive ones since the probabilistic forecast does not expect large deviations.  However, based on the historical dataset, the recursive methods determine larger amounts of reserves are required. 
The interval between $16$--$18$ h shows the opposite situation.  In this case, the recursive methods assign considerably less reserve than their anticipative counterparts.  In contrast, the anticipative methods allocate larger reserves on the potentially large deviations captured by the probabilistic forecast. 
%

\subsection{Risk Assessment}\label{sec:risk_assessment}
The previous sections assess the benefits of both the recursive and the anticipative approaches qualitatively.  However, quantitative assessment would shed light on the actual expected performance.  To carry such assessment, \textit{risk}, is used as metric,~\cite{ortega2020risk}. 

\subsubsection{Risk Quantification}
Risk ($\rho$) is the expectation of loss~\cite{wald1945statistical, degroot2005optimal}, and in this context is the loss due to insufficient reserves to accommodate the materialized deviations.  This is given as $\rho = P \times \xi$, where $P$ is the probability of undergoing deviations greater than the reserve requirement, and $\xi$ is the magnitude of such deviations.  Thus, the risk of being `short' (i.e., due to insufficient upward reserve) and the risk of being `long' (i.e., due to insufficient downward reserve) are given as, \cite{ortega2020risk}:
\begin{subequations}\label{eq:risk_short_long}
\begin{equation}
{\rho_{\rm short}} = \left( \int_{r^{\rm{up}}}^{\max(\epsilon)} \mkern-18mu f(\epsilon)\,d\epsilon \right) \times 
\left( \max(\epsilon) - r^{\rm{up}} \right),
\end{equation}
\begin{equation}
\rho_{\rm{long}} = \left( \int^{r^{\rm{dn}}}_{\min(\epsilon)} \mkern-18mu f(\epsilon)\,d\epsilon \right) \times 
\left( - r^{\rm{dn}} - \min(\epsilon) \right),
\end{equation}
\end{subequations}
where $f(\epsilon)$ is the PDF of the historical deviations, $\max(\epsilon)$ and $\min(\epsilon)$ are the ends of the positive and negative tails of the probability distribution of the deviations, respectively.

\subsubsection{Risk Assessment at Period ``t''}

To understand the differences between the recursive and anticipative reserve assessment methods, it is helpful to analyze and determine risk at specific periods of time.  For instance, consider $t=11$ and $t=16$, which have opposite reserve requirements between recursive and anticipative methods.
Figs.~\ref{fig:risk_12hrs} and~\ref{fig:risk_17hrs} show the PDFs for the chosen time intervals.  The PDFs are built using the exact historical deviations for net demand, (see eq. ~\eqref{eq:error}). 

To simplify the analysis and visualization, the deterministic-based method is used as the recursive method, and the prediction interval method is the anticipative method.  Fig.~\ref{fig:risk_12hrs} shows the PDF at $t=11$ h, where it can be seen that the downward reserve requirements are virtually the same for both approaches.  However, the upward reserve requirements are different.
The recursive approach requires larger amounts of upward reserves because there is a large probability of large positive deviations according to the historical PDF.  On the other hand, the anticipative approach is myopic to historical deviations, and thus the resulting reserve requirements are more modest.

The PDF at $t = 16$ h is shown in Fig.~\ref{fig:risk_17hrs}, and as in the previous case, the downward reserve requirements for both the recursive and anticipative methods are virtually the same.  However, the upward reserve requirements are different.  
The recursive method requires less reserve because there is a low probability of undergoing large deviations (with respect to the historical PDF).  In contrast, the anticipative method requires large amounts of reserve because in the probabilistic forecast (Fig.~\ref{fig:probabilistic_forecast}) there is a large probability of deviations from the central forecast to occur. 

However, for both $t=11$ and $t=16$ h, the hybrid method outperforms any individual approach, Fig.~\ref{fig:risk}.  This is because the hybrid method simultaneously combines the benefits of calibrating reserves with respect to the historical deviations and the probabilistic forecast. 

\subsubsection{Risk Comparison}
\begin{figure}[htb]
 \centering
  \begin{subfigure}[t]{1.72in}
     \centering
     \includegraphics[width=\textwidth]{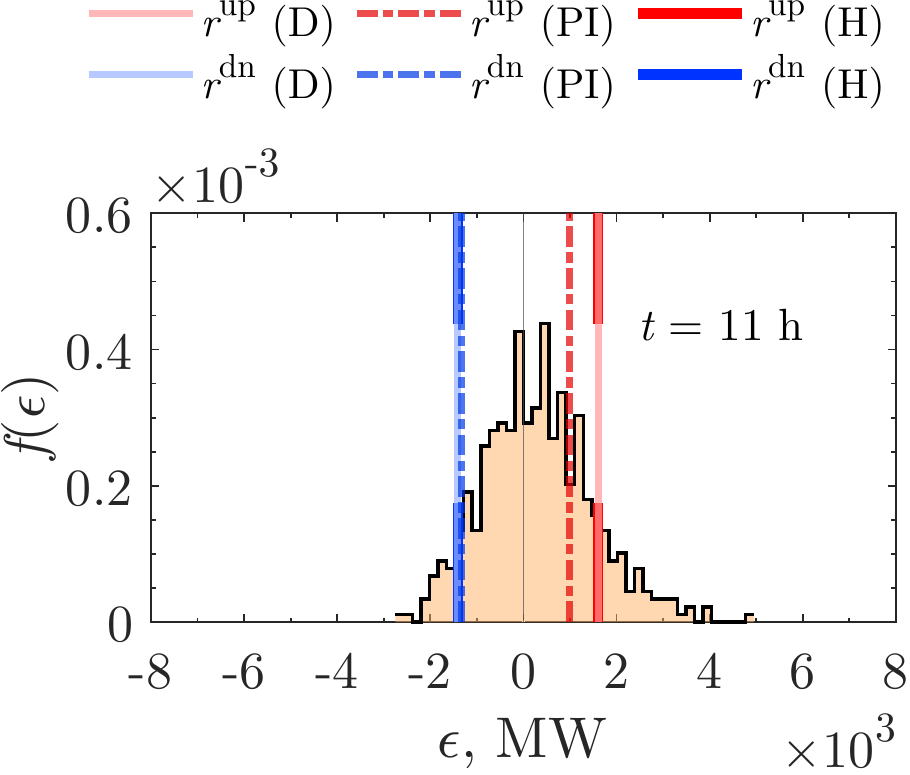}
     \caption{PDF of risk at $11$ hrs.}
     \label{fig:risk_12hrs}
 \end{subfigure}
 \hfill
 \begin{subfigure}[t]{1.72in}
     \centering
     \includegraphics[width=\textwidth]{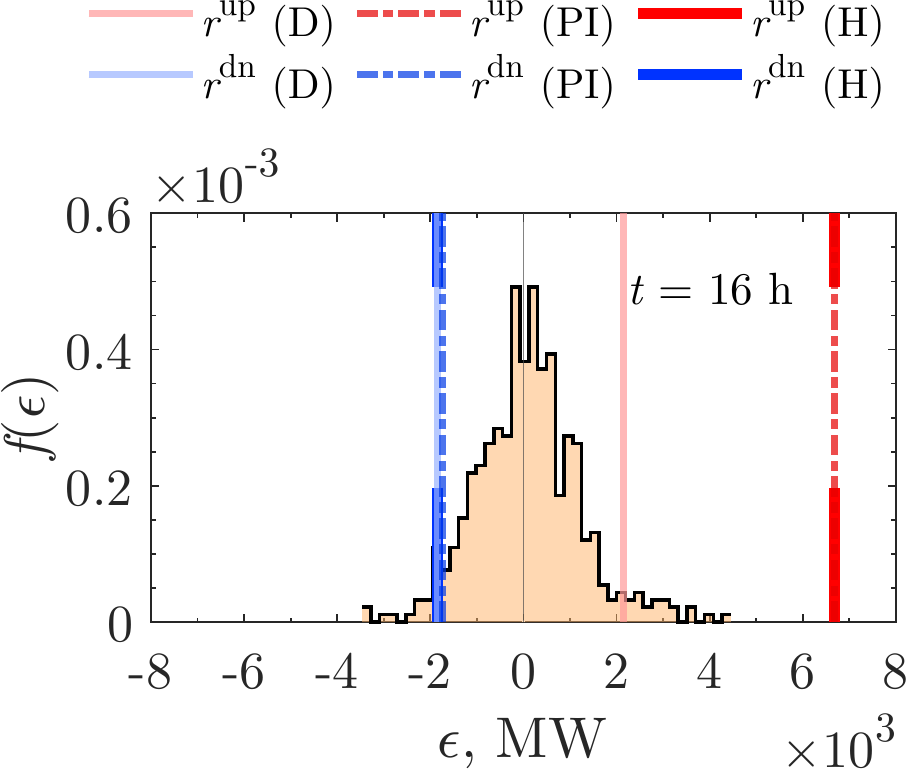}
     \caption{PDF of risk at $16$ hrs.}
     \label{fig:risk_17hrs}
 \end{subfigure}
 \begin{subfigure}[t]{1.72in}
     \centering
     \includegraphics[width=\textwidth]{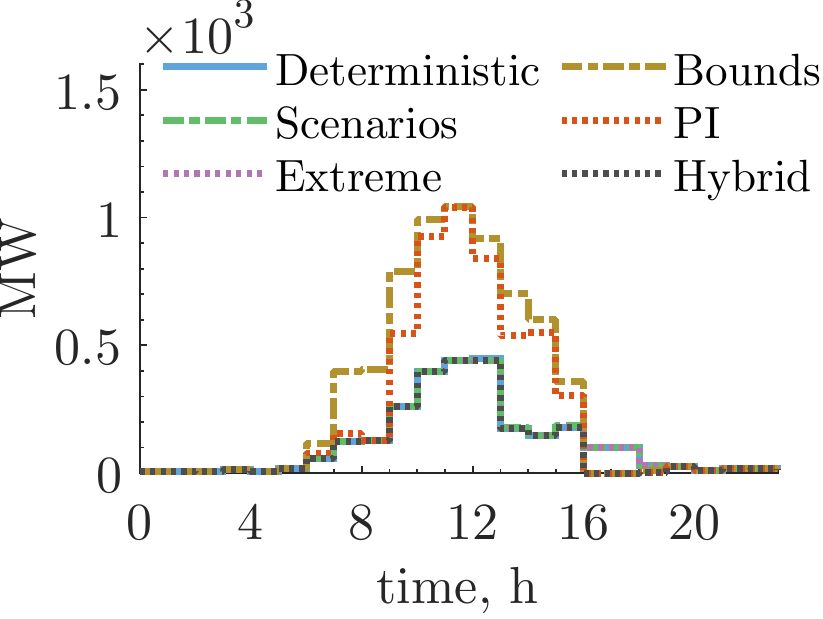}
     \caption{Risk of being short.}
     \label{fig:risk_short}
 \end{subfigure}
 \hfill
 \begin{subfigure}[t]{1.72in}
     \centering
     \includegraphics[width=\textwidth]{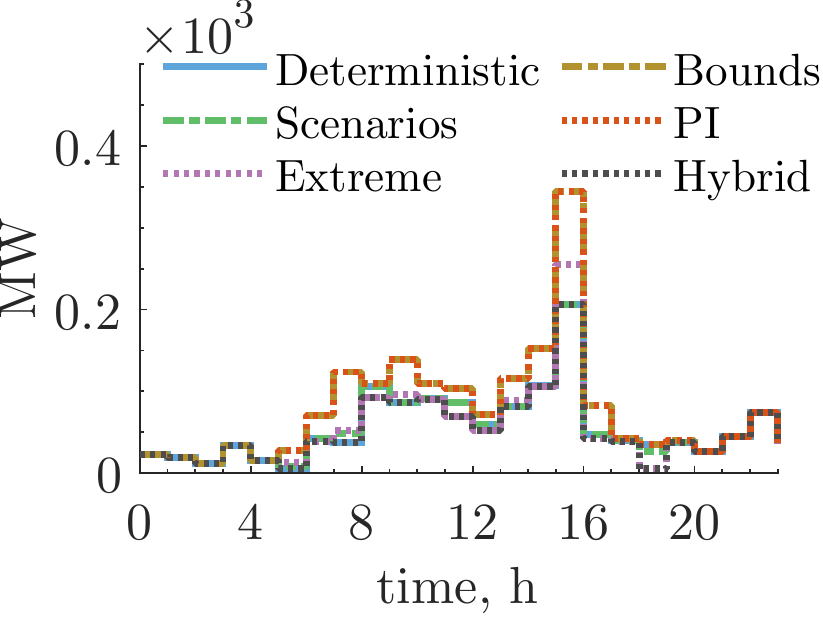}
     \caption{Risk of being long.}
     \label{fig:risk_long}
 \end{subfigure}
    \caption{Risk assessment for reserve requirements from different methods.}
    \label{fig:risk}
\end{figure}

The risk profiles for the reserves computed using the proposed methods are shown in Fig.~\ref{fig:risk}.
From this figure, it can be seen that the recursive methods result in a lower risk of being long, Fig.~\ref{fig:risk_long}.  This is not surprising because the recursive methods train with historical data and thus result in larger amounts of downward reserve than the anticipative approaches (Fig.~\ref{fig:reserve_down}) As a consequence, the system would be less likely to undergo deviations that cannot be accommodated.

This figure also shows that the recursive methods result in lower risk than the anticipative ones between $6$--$15$ h.  Again, this is because the recursive methods expect larger deviations than those captured by the probabilistic forecast and thus schedule larger reserve amounts.  In contrast, between $16$--$18$ h, the anticipative methods result in lower risk than the recursive approaches, Fig.~\ref{fig:risk_short}.  This is because the anticipative methods expect large deviations during this period, which have not been captured (at least frequently) in the historical dataset.  Therefore these methods increase the reserve requirements.

While there is a trade-off between recursive and anticipative approaches, the hybrid methods schedule reserves based on the combination of both.  As a result, the system is protected against deviations whether they are captured in the historical dataset, on the probabilistic forecast, or on both.  Therefore the risk profile of the hybrid approach is the lowest at all times, Fig.~\ref{fig:risk_short}.
%

\subsection{Sensitivity Analysis}
There are several parameters that could be adjusted when using the different approaches.  Therefore it is essential to understand their impact on the resulting reserve requirements. 
Specifically, the effect of the confidence interval `${\rm CI}$' on the recursive approaches based on~\eqref{eq:reserve_requirements}, and the impact of the prediction interval `PI' on the anticipative methods are discussed. 

\begin{figure}[htb]
 \centering
 \begin{subfigure}[t]{1.72in}
     \centering
     \includegraphics[width=\textwidth]{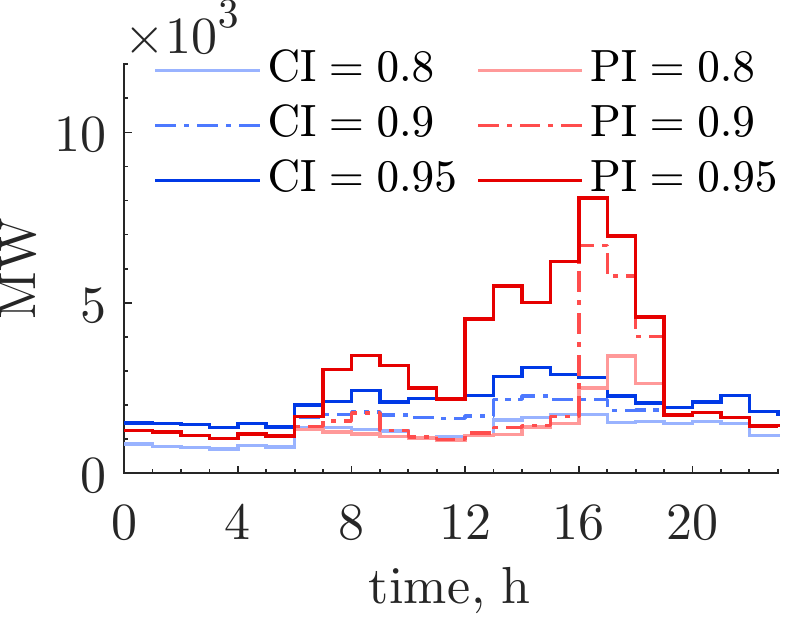}
     \caption{Reserve up.}
     \label{fig:sensitivity_up}
 \end{subfigure}
 \hfill
 \begin{subfigure}[t]{1.72in}
     \centering
     \includegraphics[width=\textwidth]{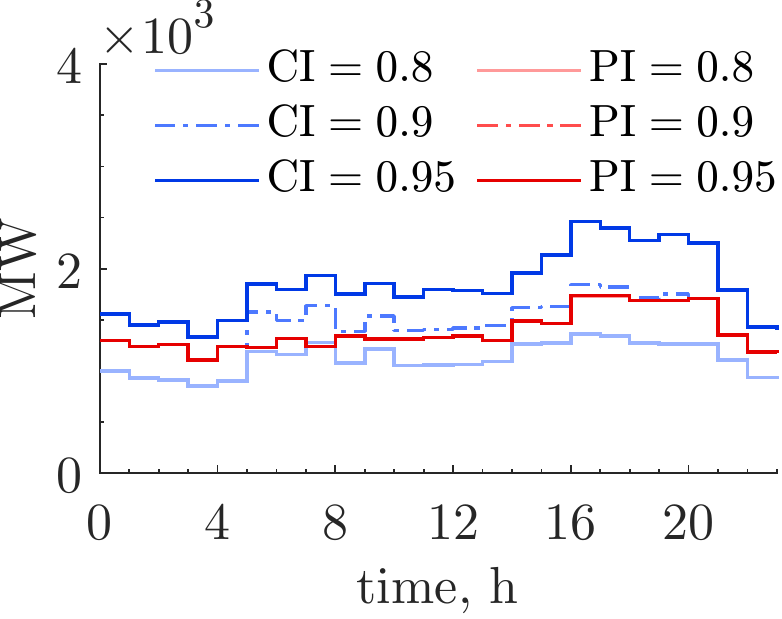}
     \caption{Reserve down.}
     \label{fig:sensitivity_down}
 \end{subfigure}
    \caption{Sensitivity of upward and downward reserves for recursive and anticipative methods.}
    \label{fig:sensitivity_analysis}
\end{figure}

Again, for the sake of simplicity and presentation, only a recursive and an anticipative method are selected.  For the recursive methods, the deterministic method is chosen, (eq.~\eqref{eq:reserve_requirements}), and the reserve requirements are computed for ${\rm CI} = 0.8, 0.9, 0.95$.  For the anticipative methods, the prediction interval method is used, (eq.~\eqref{eq:reserve_PI}).  In this case, there are two parameters to adjust: `PI' for the solar reserve components and `$\rm{CI}$' for the load and wind production reserve components.  Since for the anticipative approach only the solar uncertainty sensitivity is of interest, the confidence interval for load and wind is set as ${\rm{CI}} = 0.9$; and solar reserve requirements are computed for ${\rm PI} = 0.8, 0.9, 0.95$.
The results are shown in Fig.~\ref{fig:sensitivity_analysis}.

The sensitivity of upward reserves is shown in Fig.~\ref{fig:sensitivity_up}.  In this figure, it can be seen that the reserve requirements increase for the recursive method (in blue) as ${\rm CI}$ increases.  In contrast, for the anticipative method (in red), the upward reserve requirements increase more rapidly as PI increases.  This effect in the anticipative method is a direct consequence of the probabilistic forecast; the most sensitive time interval is between $16$ and $18$ h for the probabilistic forecast (Fig.~\ref{fig:probabilistic_forecast}) since it has a large probability spread for solar deviations below the central forecast. 

Fig.~\ref{fig:sensitivity_down} shows the sensitivity of downward reserves. In this figure, it can be seen that the reserve requirements increase for the recursive method as $\rm{CI}$ increases.  However, the downward reserves for the anticipative method are insensible to the $\rm{PI}$ parameter.  This is because, for this particular day, the probabilistic forecast has an extremely narrow uncertainty envelope above the central forecast. 
As a result, the solar reserve requirements are virtually insensitive to the $\rm PI$.
%

\subsection{Risk-based and Hybrid Methods}

Section~\ref{sec:risk_assessment} uses \textit{risk} as a metric.  However, risk can also be enforced as a ceiling/limit ($\rho_{\rm limit}$.) to determine the reserve requirements,~\cite{ortega2020risk}.  That is, the reserves are computed such that the risk at all periods is equal or lower than the risk ceiling ($\rho_{\rm short/long} \leq \rho_{\rm limit}$). 

\begin{figure}[htb]
 \centering
  \begin{subfigure}[t]{1.72in}
     \centering
     \includegraphics[width=\textwidth]{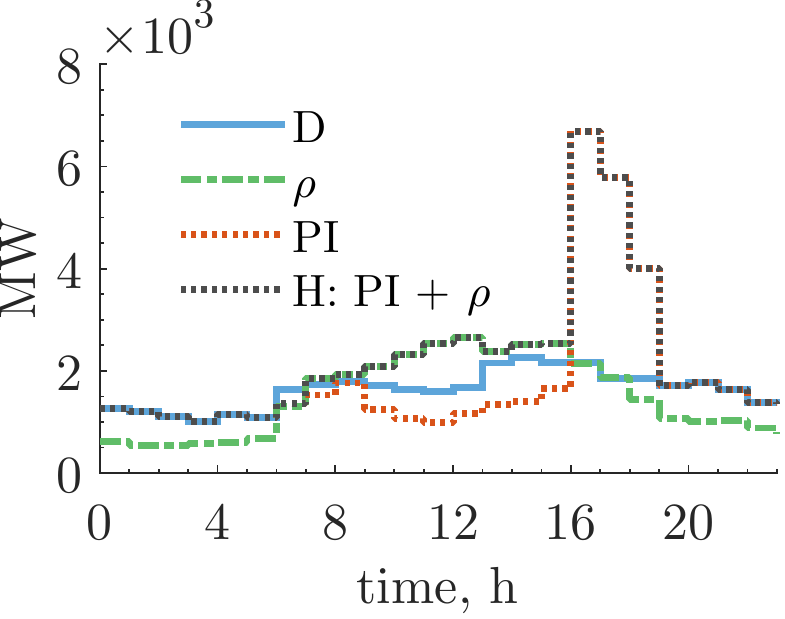}
     \caption{Reserve up.}
     \label{fig:r_up_risk}
 \end{subfigure}
 \hfill
 \begin{subfigure}[t]{1.72in}
     \centering
     \includegraphics[width=\textwidth]{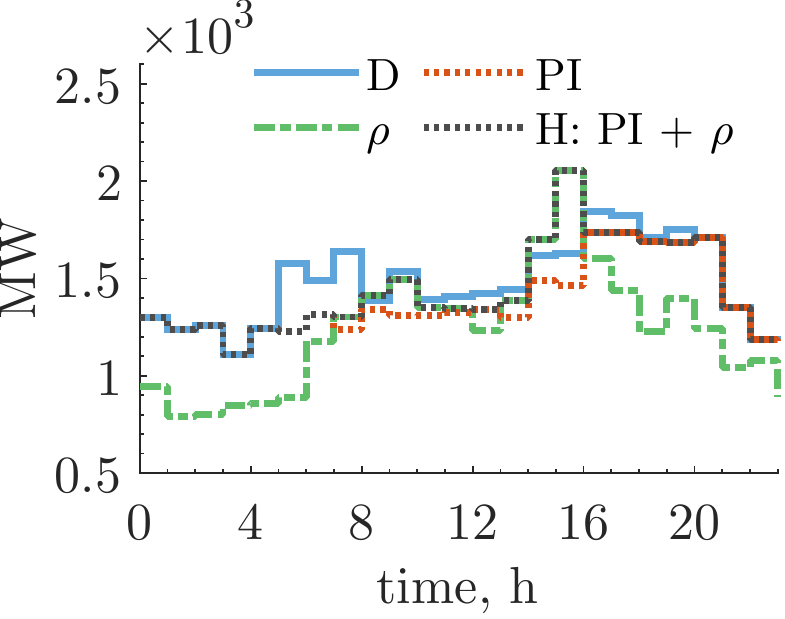}
     \caption{Reserve down.}
     \label{fig:r_dn_risk}
 \end{subfigure}
 \begin{subfigure}[t]{1.72in}
     \centering
     \includegraphics[width=\textwidth]{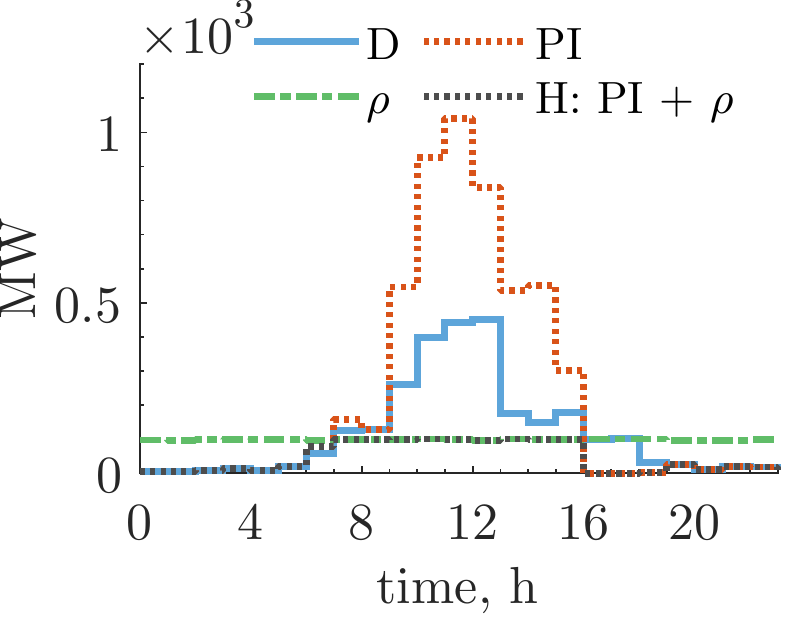}
     \caption{Risk of being short.}
     \label{fig:risk_short_risk}
 \end{subfigure}
 \hfill
 \begin{subfigure}[t]{1.72in}
     \centering
     \includegraphics[width=\textwidth]{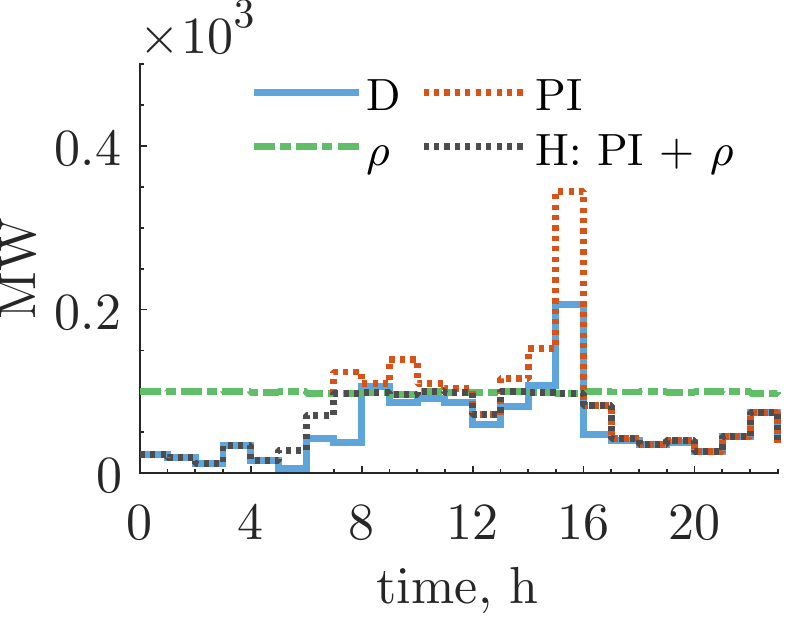}
     \caption{Risk of being long.}
     \label{fig:risk_long_risk}
 \end{subfigure}
    \caption{Comparison of reserve requirements and risk assessment for risk-based, recursive, anticipative, and hybrid methods.}
    \label{fig:risk-based}
\end{figure}

Risk-based reserve requirements can also be used as part of the reserve scheduling process in hybrid methods.  In this case, for the sake of clarity, a recursive (deterministic), anticipative (prediction interval, $\rm PI$), and hybrid (${\rm PI} + \rho$) methods are compared in Fig.~\ref{fig:risk-based}. 
In this figure, it can be seen that the risk-based method produces reserves that are equal or below the limit of $100\,{\rm MW}$ at all periods.  However, risk-based reserve requirements do not capture the large event the need for larger reserves as the anticipative method (hours $16$--$18$).  This is because the risk-based method is recursive, as it determines the requirements based excursively on the distributions of the historical deviations.  Combining the risk-based method with an anticipative approach hedges against historical deviations from the risk perspective, and it also protects against the expected deviations captured in the probabilistic forecast, Figs.~\ref{fig:r_up_risk} and ~\ref{fig:risk_short_risk}.  
%

\section{Conclusion}\label{sec:conclusion}
This work proposes different methods to determine operating reserve requirements from probabilistic forecasts.  The methods can be divided into recursive, anticipative, and hybrid groups.  The recursive approaches set the reserves based on registered deviations on historical datasets.  The anticipative approaches rely more heavily on the shape and properties of the probabilistic forecast.  The hybrid methods combine the benefits of both the recursive and the anticipative approaches. 

These approaches give more insight for better reserve procurement by not only considering the most likely materialization but also regarding the specific uncertainty of a given day.  A thorough analysis was conducted using actual data from CAISO and quantifying the associated risk with each reserve determination method.  The hybrid approach outperformed any individual method because it combines the benefits of both the historical and probabilistic forecast information. 

Utilities and ISOs currently rely on rules-of-thumb, deterministic, static, and manual incorporation of probabilistic forecasts when utilizing historical data approaches to determine reserves.  The ability to systematically incorporate probabilistic forecasts on the reserve dimensioning will allow them to protect the system against less frequent but potentially more damaging conditions such as adverse weather. 

The presented approaches integrate seamlessly with the workflow of balancing authorities. In particular, hybrid methods would allow them to transition from methods based on deterministic forecasts to fully probabilistic methods. Or retain the benefits of the deterministic rules while also attaining the benefits of the probabilistic approaches when these detect potentially risky conditions.

In this work, the benefit of the approaches has been presented qualitatively in terms of shapes and statistical properties and qualitatively in terms of system risk.  Ongoing work is expected to demonstrate these benefits of probabilistic approaches using production cost model simulations for the Hawaiian Electric, Duke, and Southern Company systems. 


\bibliographystyle{IEEEtran}
\bibliography{bib/bibliography.bib,bib/probabilistic_forecast.bib}

\end{document}